\documentclass{article}
\usepackage{latexsym,bm,amsmath}
\usepackage{amsfonts}
\usepackage{amsfonts}
\usepackage{amsfonts}
\usepackage{amsmath}
\usepackage{amssymb}
\usepackage{mathptmx}   
\usepackage{times}      
\usepackage{amsmath}    
\usepackage{graphicx}
\usepackage{subfigure}
\begin{document}
\hoffset = -1truecm \voffset = -2truecm
\title {\bfseries A Riemann-Hilbert Approach to the Complex Sharma-Tasso-Olver Equation on the Half Line }
\author{\bfseries Ning Zhang $^{\dag a,b}$,Tiecheng Xia $^{a}$, Beibei Hu$^{a,c}$ \\
\small $^a$ Department of Mathematics, Shanghai
University, Shanghai, 200444, PR China\\
\small $^b$ Department of Basical Courses, Shandong University of Science and Technology, Taian 271019, PR China\\
\small $^c$ School of Mathematicas and Finance, Chuzhou University, Anhui, 239000, China}
\date{}
\footnotetext{$^\dag$: Corresponding author.\\
\indent Email addresses : zhangningsdust@126.com(N.Zhang);  xiatc@shu.edu.cn(T.c.Xia); hu\_chzu@163.com(B.b.Hu)}
\maketitle {\setlength{\baselineskip}{15pt} {\bfseries \noindent
Abstract:} In this paper, we use the Fokas method to analyze  the complex Sharma-Tasso-Olver(cSTO) equation $u_y+\frac{1}{2}(u_{xxx}-3i(uu_x)_x-3u^2u_x)=0$
 on the half line $0<x<\infty$. Assuming that   the solution $u(x,y)$ of the cSTO equation  is exists, we show that it can be represented in terms of the solution of a matrix Riemann-Hilbert problem(RHP) formulated in the complex $\lambda-$plane (the Fourier plane), which has a jump matrix with explicit $(x,y)-$dependence involving four scalar functions of $\lambda$, called spectral
functions. The spectral functions $\{a(\lambda),b(\lambda)\}$ and $\{A(\lambda),B(\lambda)\}$ are obtained from the initial data $u_0(x)=u(x,0)$ and the boundary data $g_0(y)=u(0,y),g_1(y)=u_x(0,y),g_2(y)=u_{xx}(0,y)$, respectively. The problem has the jump across $\{Im \lambda^6=0 \}$. The spectral functions are not independent,
but related by a compatibility condition, the so-called global relation. Given initial and boundary values
$ u_0( x ), g_ 0 ( y ), g_ 1 ( y )$ and $g_2(y)$ such that there exist spectral functions satisfying the global relation, we show
that the function $u ( x , y ) $ defined by the above Riemann-Hilbert problem exists globally and solves the cSTO
 equation with the prescribed initial and boundary values.
\\
{\bfseries \noindent 2010 MR Subject Classification 35G31; 35Q15}
\section{ Introduction } \vskip0.5cm
\indent ~~~~A unified method for analyzing boundary value problems with decaying initial data, extending ideas of the so-called
inverse scattering transform(IST) method, was discovered in 1967[1].  This
method can be thought of as a nonlinear Fourier transform(FT) method. However, this nonlinear FT is not the same for every nonlinear evolution equation, but it
is constructed from the $x$ part of the Lax pair. Furthermore, neither the direct nonlinear
FT of the initial data, nor the inverse nonlinear FT can be
expressed in closed form: the former involves a linear Volterra integral equation and
the latter involves a matrix Riemann-Hilbert problem(RHP).  The $y$ part is used only to determine the evolution of the direct nonlinear
FT[2,3]. However, in many laboratory and field situations, the wave motion is initiated by
what corresponds to the imposition of boundary conditions rather than initial conditions. This naturally leads to the formulation of an initial-boundary value (IBV) problem instead of a pure initial value problem.\\
\indent In 1997, a new unified approach based on the Riemann-Hilbert factorization problem
to solve IVB problems for linear and nonlinear integrable PDEs was presented by Fokas [4], we call that Fokas method. The Fokas method provides a generalization of the inverse scattering formalism from initial value to IVB problems, it is not only able to identify the linearizable class of boundary conditions, but what is more important, it is able to solve this class of boundary value problems as effectively as the usual class of initial value problems on the line. Over the last almost two decades,  it was further developed by Its[5,6], de Monvel[7,8], Lenells[9-11], Engui Fan and JianXu[12-14] ect. After intense investigation it appears that there exists now an elegant, rigorous method for solving the half line problem for integrable nonlinear PDE＊s. An effort has been made to present this new method in a form that will be accessible to a wide audience. This is important, since it is hoped that researchers will consider using this method to solve a large class of physically important boundary value problems which remain open.  Furthermore, it is interesting
that this new method gives rise to a new numerical method for solving linear elliptic boundary value problems (this method is based on the numerical solution of the global relation [15]).\\
\indent
The Sharma-Tasso-Olver(STO) equation
\begin{equation}\numberwithin{equation}{section}
u_y+\frac{1}{2}(u_{xxx}-3(u_xu)_x-3u^2u_x)=0
\end{equation}
was first derived as an example of odd members of Burgers hierarchy by extending the "linearization" achieved through the Cole-Hopf ansatz to equations containing as highest
derivatives odd space derivatives[16]. In recent years, many physicists and mathematicians have paid much attention to the STO equation  due to its importance in mathematical physics.
The bi-Hamiltonian formulation and the generalized Poisson bracket are pre-
sented in [17], the exact traveling solutions and symmetries of the STO equation were extensively studied in the literature[17-19].  Moreover, the multiple solitons, exact traveling solutions, kinks
solutions, non-traveling wave solutions of the STO equation were obtained with different approaches[18-22].
In this paper, we  concerned with the cSTO equation derived by Fan[23]
\begin{equation}\numberwithin{equation}{section}
 u_y+\frac{1}{2}(u_{xxx}-3i(u_xu)_x-3u^2u_x)=0.
\end{equation}
If we let $i\lambda\rightarrow\lambda$ in the spectral problem (1.2), then we will get the real version of the STO equation (1.1).
 The purpose of this paper is to  analyze the IBV problem for the cSTO equation (1.2) on the half line using Fokas method.
 That is, in the  quarter $(x,y)-$plane
 \begin{equation}
\Omega=\{(x,y)|0\leq x<\infty,0\leq y<L\},
\end{equation}\
where $L$ is a positive finite constant. The side $\{y=L, 0<x<\infty\}$, $\{x=0, 0<y<L\}$ and $\{y=0, 0<x<\infty\}$ will be referred to as side (1), (2) and (3), respectively, see Fig.1.\\

\begin{figure}[h]
\centering
\includegraphics[width=3.0in,height=2.5in]{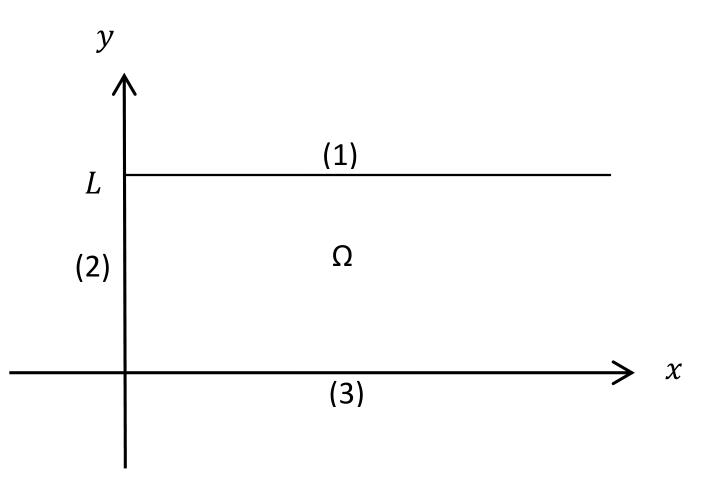}
\caption{The $(x,y)$-domaint.}
\label{fig:graph}
\end{figure}

 Assuming that the solution $u(x,y)$ of the cSTO equation exists, define
 \begin{equation}
 Initial~~ value:u(x,0)=u_0(x),~~~~~0\leq x<\infty,
\end{equation}
 and
  \begin{equation}
 Boundary~~ value:u(0,y)=g_0(y), u_{x}(0,y)=g_1(y), u_{xx}(0,t)=g_2(y)~~~0\leq y<\infty.
\end{equation}
We show that it can be represented in terms of the solution of a matrix RHP formulated in the plane of the complex spectral parameter $\lambda$.
The jump matrix has explicit $(x,y)$ dependence and is given in terms of the spectral functions $\{a(\lambda),b(\lambda)\}$ and $\{A(\lambda),B(\lambda)\}$, which are obtained from the initial data (1.4) and the boundary data (1.5), respectively. The problem has the jump across $\{Im \lambda^6=0 \}$. The spectral functions are not independent,
but related by a compatibility condition, the so-called global relation, which is an algebraic equation coupling $\{a(\lambda),b(\lambda)\}$ and $\{A(\lambda),B(\lambda)\}$.
An important advantage of the methodology of ref.[4] is that it yields precise information about the long time asymptotic of the solution.
The usefulness of the asymptotic information for a physical problem can be understood by considering an example where the field is at rest at some initial time $y=0$. Assuming that we can create or measure the waves emanating from some fixed point, say $x=0$. In space, we arrive at an initial-boundary value problem for which the initial datum $u_0(x)$ vanishes identically, while the boundary values $g_0(y)$, $g_1(y)$, $g_2(y)$ are at our disposal. An analysis of the asymptotic behavior of the solution then provides information about the long time effect of the known boundary values. In \textbf{Section 2}, we give some summary results and the basic RHP of the cSTO equation (1.2). In \textbf{Section 3}, we construct the inverse spectral mappings for the initial and boundary
values. In the following \textbf{Section 4}, the spectral functions $a(\lambda),b(\lambda)$ and $A(\lambda),B(\lambda)$ are investigated and the  principal RHP is presented. \\

\section{ The Basic RHP of CSTO Equation}
\indent~~~ Frist, we explain what the RHP is.\\
\textbf{Definition 2.1.} Let the contour $\Gamma$ be the union of a finite number of smooth and oriented curves on the Riemann sphere $\mathbb{C}$, such that $\mathbb{C}\setminus \Gamma$ has only a finite number of connected components. Let $J(\lambda)$ be a $2\times2$ matrix defined on the contour $\Gamma$, the RHP $(\Gamma,J)$ is the problem of finding a $2\times2$ matrix-valued function $M(\lambda)$ that satisfies
\begin{description}
  \item[(i)] $M(\lambda)$ is analytic for all $\lambda\in \mathbb{C}\setminus \Gamma$, and extends continuously to the contour $\Gamma$;
  \item[(ii)] $M_{+}(\lambda)=M_{-}(\lambda)J(\lambda), \lambda\in \Gamma$;
  \item[(iii)] $M(\lambda)\rightarrow I,$ as $\lambda\rightarrow\infty$.
\end{description}
\textbf{\textbf{2.1.}Lax pair for the cSTO equation}\\
\indent The cSTO equation (1.2) admits the Lax pair[23]
\begin{equation}
\left\{
  \begin{array}{ll}
    \psi_x(x,y;\lambda)=U(x,y;\lambda)\psi(x,y;\lambda), \\
    \psi_y(x,y;\lambda)=V(x,y;\lambda)\psi(x,y;\lambda),
  \end{array}
\right.
\end{equation}
where
\begin{equation} U(x,y;\lambda)=\left(
   \begin{array}{cc}
     -i\lambda^2-\frac{1}{2}iu & \lambda u \\
     2\lambda & i\lambda^2+\frac{1}{2}iu \\
   \end{array}
 \right),
 V(x,y;\lambda)=\left(
   \begin{array}{cc}
     V_{11} & V_{12} \\
     V_{21} & -V_{11} \\
   \end{array}
 \right),
\end{equation}
and
$V_{11}=-2i\lambda^6-2i\lambda^4u+\lambda^2(u_x-iu^2)+\frac{i}{4}u_{xx}+\frac{3}{4}uu_x-\frac{i}{4}u^3,$
$V_{12}=2\lambda^5u+\lambda^3(iu_x+u^2)+\lambda(-\frac{1}{2}u_{xx}+\frac{3}{2}iuu_x+\frac{1}{2}u^3),$
$V_{21}=4\lambda^5+2\lambda^3u+\lambda(u_{x}+u^2).$
Indeed, one can check that the compatibility condition of the two equations in (2.1) yields the zero-curvature equation $U_y-V_x+[U,V] = 0$,
which is exactly the cSTO equation (1.2). Here the square bracket is the usual matrix commutator defined as $[U,V]=UV-VU$.\\
\indent In what follows, rather than work with the original form of Lax pair(2.1), it turns out to be more convenient by introducing
$$P=\left(
      \begin{array}{cc}
        0 & u \\
        2 & 0 \\
      \end{array}
    \right),
\sigma_{+}=\left(\begin{array}{cc}0 & 1 \\0 & 0 \\\end{array}\right),\sigma_{-}=\left(\begin{array}{cc}0 & 0 \\1 & 0 \\\end{array}\right),
\sigma_{3}=\left(\begin{array}{cc}1 & 0 \\0 & -1 \\\end{array}\right),$$
$\sigma_{3}$ denotes the third Pauli's matrix, we can rewrite the Lax pair (2.1) in a matrix form
\begin{equation}
\left\{
  \begin{array}{ll}
    \psi_x+i\lambda^2\sigma_{3}\psi=(-\frac{1}{2}iu\sigma_{3}+\lambda P)\psi, \\
    \psi_y+2i\lambda^6\sigma_{3}\psi=(A+B\sigma_{3})\psi,
  \end{array}
\right.
\end{equation}
where $A=2\lambda^5P+\lambda^3(iP_x+uP)+\lambda(-\frac{1}{2}P_{xx}+\frac{3}{8}i(P^3)_x-\frac{1}{2}iPP_x\sigma_-+\frac{1}{2}u^2P),B=-2i\lambda^4u+\lambda^2(u_x-iu^2)+\frac{1}{4}iu_{xx}+\frac{3}{4}uu_x-\frac{1}{4}iu^3$.
Extending the column vector $\psi$ to a $2\times 2$ matrix and letting
\begin{equation}
\Psi=\psi e^{i(\lambda^2x+2\lambda^6y) \sigma_3}
\end{equation}
Using Equ.(2.4), the original form of Lax pair (2.1) can be rewritten in the form
\begin{equation}
\left\{
  \begin{array}{ll}
    \Psi_x+i\lambda^2[\sigma_{3},\Psi]=(-\frac{1}{2}iu\sigma_{3}+\lambda P)\Psi, \\
    \Psi_t+2i\lambda^6[\sigma_{3},\Psi]=(A+B\sigma_{3})\Psi,
  \end{array}
\right.
\end{equation}
which can be written in full derivative form
\begin{equation}
d(e^{i(\lambda^2x+2\lambda^6y )\hat{\sigma}_3}\Psi(x,y;\lambda))=e^{i(\lambda^2x+2\lambda^6y )\hat{\sigma}_3}Q(x,y;\lambda)\Psi,
\end{equation}
where, $\hat{\sigma}_{3}$ denotes the matrix commutator with $\sigma_{3}$, $\hat{\sigma}_{3}M=[\sigma_{3},M]$, then $e^{\hat{\sigma}_{3}}$ can be easily computed  $e^{\hat{\sigma}_{3}}M=e^{\sigma_{3}}Me^{-\sigma_{3}}$( $M$ is a $2\times2$ matrix), $Q(x,y;\lambda)=Q_1(x,y;\lambda)dx+Q_2(x,y;\lambda)dy$, and $Q_1(x,y;\lambda)=-\frac{1}{2}iu\sigma_{3}+\lambda P, Q_2(x,y;\lambda)=A+B\sigma_{3}$.\\
$\textbf{Statement of the problem.}$ Let $u(x,y)$satisfy a nonlinear evolution equation with spatial
derivatives of order $n$, on the half line $0 < x < \infty$, and for $ 0 < y < L$, where $L$
is a positive constant. Let $u(x,y)$ satisfy decaying initial conditions at $y = 0$, as well
as appropriate boundary conditions at $x = 0$. Assume that this PDE admits a Lax pair
formulation of the type (2.5), i.e. assume that this PDE is equivalent to the compatibility
condition of Eq.(2.2). Then such an initial-boundary value problem can be analyzed
using the Fokas and Lenells method[10,14,24,25].
\textbf{\textbf{2.2.} Spectral Analysis and Asymptotic Analysis}\\
\indent Consider that a solution of Eq.(2.5) is of the form \\
\begin{equation}
\Psi(x,y;\lambda)=D+\sum_{j=1}^5\frac{\Psi_j(x,y;\lambda)}{\lambda^j}+O(\frac{1}{\lambda^6}), \,\,\,\,\,\lambda\rightarrow\infty,
\end{equation}
where $D,\Psi_j(j=1,2,\cdots 5.)$ are independent of $\lambda$. Substituting the above expansion into the first equation of (2.5), and comparing the same order of frequency of $\lambda$, we have\\
\begin{equation}\begin{array}{ll}
o(\lambda^2):i[\sigma_3,D]=0 ,\\
o(\lambda):i[\sigma_3,\Psi_1]=PD,\\
o(1):D_{x}=\frac{i}{4}P^2\sigma_3D.
\end{array}
\end{equation}
Let $M^{(o)}$ denotes the off-diagonal part of $M$, $M^{(d)}$ denotes the diagonal part of $M$, $M$ is a $2\times2$ matrix.
From (2.8) we have
\begin{equation}
D=\left(\begin{array}{cc}
      D^{11} & 0 \\
              0 & D^{22} \\
           \end{array}
\right),\Psi_1^{(o)}=\frac{i}{2}PD\sigma_3ㄛ
\end{equation}
and
\begin{equation}
D_x=\frac{i}{2}u\sigma_3D,
\end{equation}
where, $D$ is a diagonal matrix.\\
On the other hand, substituting the above expansion into the second equation of (2.5), we have
\begin{equation}\begin{array}{ll}
o(\lambda^6):2i[\sigma_3,D]=0,\\
o(\lambda^5):2i[\sigma_3,\Psi_1]=2PD,\\
o(\lambda^4):2i[\sigma_3,\Psi_2]=-2iuD \sigma_3+2P\Psi_1,\\
o(\lambda^3):2i[\sigma_3,\Psi_3]=(iP_x+uP)D-2iu\sigma_3\Psi_1+2\Psi_2,\\
o(\lambda^2):2i[\sigma_3,\Psi_4]=(u_x-iu^2)\sigma_3D+(iP_x+uP)\Psi_1-2iu\sigma_3\Psi_2+2P\Psi_3,\\
o(\lambda^1):2i[\sigma_3,\Psi_5]=(-\frac{1}{2}P_{xx}+\frac{3i}{8}(P^3)_x-\frac{i}{2}PP_x\sigma_-+\frac{1}{2}u^2P)D+(u_x-iu^2)\sigma_3\Psi_1\\
~~~~~~~~~~~~~~~~~~~~~~~~~~~~~~~~~~+(iP_x+uP)\Psi_2-2iu\sigma_3\Psi_3+2P\Psi_4,\\
o(1):D_{y}=(\frac{i}{4}u_{xx}+\frac{3}{4}uu_x-\frac{i}{4}u^3)\sigma_3D+(-\frac{1}{2}p_{xx}+\frac{3i}{8}(P_x)^3-\frac{i}{2}PP_x\sigma_-+\frac{1}{2}u^2P)\Psi_1\\
~~~~~~~~~~~~~~~~~~+(u_x-iu^2)\sigma_3\Psi_2+(iP_x+uP)\Psi_3-2iu\sigma_3\Psi_4+2P\Psi_5.
\end{array}
\end{equation}

From (2.11), we have
\begin{equation}
\begin{array}{ll}
\Psi_2^{(o)}=\frac{1}{2}iP\Psi_1^{(d)},\\
\Psi_3^{(o)}=\frac{i}{4}(iP_x+uP)D\sigma_3-\frac{1}{2}u\Psi_1^{(o)}-\frac{1}{2}iP\Psi_2^{(d)}\sigma_3,\\
\Psi_4^{(o)}=\frac{i}{4}(u_x-iu^2)D-\frac{i}{4}(iP_x+uP)\Psi_1^{(d)}\sigma_3-\frac{1}{2}u\Psi_2^{(o)}-\frac{i}{2}P\Psi_3^{(d)}\sigma_3,\\
\Psi_5^{(o)}=\frac{i}{4}(-\frac{1}{2}P_{xx}+\frac{3}{8}i(P^3)_x-\frac{i}{2}PP_x\sigma_-+\frac{1}{2}u^2P)D\sigma_3-\frac{i}{4}(u_x-iu^2)\Psi_1^{(o)}-\frac{i}{4}(iP_x+uP)\Psi_2^{(d)}\sigma_3\\
~~~~~~~~~~~~~~-\frac{1}{2}P\Psi_3^{(o)}-\frac{i}{2}P\Psi_4^{(d)}.
\end{array}
\end{equation}
and
\begin{equation}
D_y=(-\frac{1}{2}u_{xx}-\frac{3}{2}iuu_x+\frac{1}{2}u^3)_x.
\end{equation}
We note that the Eq.(1.2) admits the conservation law
\begin{equation}
u_y=(-\frac{1}{2}u_{xx}+\frac{3}{2}iuu_x+\frac{1}{2}u^3)_x.
\end{equation}
 Then the two Eq.(2.10) and Eq.(2.13) for $D$ are consistent and are both satisfied if we define
\begin{equation}
D(x,y)=exp(i\int_{(x_0,y_0)}^{(x,y)}\Delta\sigma_3),
\end{equation}
where $\Delta$ is the closed real-valued one-form, and
\begin{equation}
\begin{array}{ll}
\Delta(x,y)=\Delta_1(x,y)dx+\Delta_2(x,y) dy,\\
\Delta_1(x,y)=\frac{1}{2}u, \,\,\,\Delta_2(x,y)=-\frac{1}{4}u_{xx}+\frac{3}{4}iuu_x+\frac{1}{4}u^3,\,\,\,\ (x_0,y_0)\in \Omega.
\end{array}
\end{equation}
 simultaneity, for the convenience of calculation we denote $(x_0,y_0)=(0,0).$\\
\textbf{\textbf{2.3.} Eigenfunctions and Their Relations}\\
\indent Noting that the integral in Eq.(2.15) is independent of
the path of integration and the $\Delta(x,y)$ is independent of $\lambda$, then we can introduce eigenfunction $\mu(x,y;\lambda)$ as follows
\begin{equation}
\Psi(x,y;\lambda)=e^{i\int_{(0,0)}^{(x,y)}\Delta\hat{\sigma}_3}\mu(x,y;\lambda)D(x,y),\,\,\,\, 0<x<\infty,\,\,\,\,\,\, 0<y<L,
\end{equation}
where $\mu(x,y,\lambda)$ is a $2\times2$ matrix-valued function.
Through direct calculation, the Lax pair of Eq.(2.6) becomes
\begin{equation}
d(e^{i(\lambda^2x+2\lambda^6y)\hat{\sigma}_3}\mu(x,y;\lambda))=W(x,y;\lambda),\,\,\,\,\lambda\in\mathbb{C},
\end{equation}
where
\begin{equation}
W(x,y;\lambda)=e^{i(\lambda^2x+2\lambda^6y)\hat{\sigma}_3}N(x,y;\lambda)\mu(x,y;\lambda),
\end{equation}
\begin{equation}
N(x,y;\lambda)=N_1(x,y;\lambda)dx+N_2(x,y;\lambda)dy
\end{equation}
\begin{equation}
N_1(x,y;\lambda)=\left(
                   \begin{array}{cc}
                     -iu & \lambda u e^{-2i\int_{(0,0)}^{(x,y)}\Delta} \\
                     2\lambda  e^{2i\int_{(0,0)}^{(x,y)}\Delta} & iu \\
                   \end{array}
                 \right),
N_2(x,y;\lambda)=\left(
                   \begin{array}{cc}
                     N_2^{11} & N_2^{12} \\
                     N_2^{21} &-N_2^{11} \\
                   \end{array}
                 \right),
\end{equation}
and
\begin{equation}
\begin{array}{ll}
N_2^{11}=-2i\lambda^4u+\lambda^2(u_x-iu^2)+\frac{i}{2}u_{xx}+\frac{3}{2}uu_x-\frac{i}{2}u^3,\\
N_2^{12}=(2\lambda^5u+\lambda^3(iu_x+u^2)+\lambda(-\frac{1}{2}u_{xx}+\frac{3}{2}iuu_x+\frac{1}{2}u^3) e^{-2i\int_{(0,0)}^{(x,y)}\Delta},\\
N_2^{21}=(4\lambda^5+2\lambda^3u+\lambda(iu_x+u^2) e^{2i\int_{(0,0)}^{(x,y)}\Delta}.
\end{array}
\end{equation}
Then Eq.(2.19) for $\mu(x,y;\lambda)$ can be written as
\begin{equation}
\left\{
  \begin{array}{ll}
    \mu_x+i\lambda^2[\sigma_3,\mu]=N_1\mu,  \\
    \mu_y+2i\lambda^6[\sigma_3,\mu]=N_2\mu,
  \end{array}
\right.
\end{equation}
where $(x,y)\in \Omega,\,\,\,\lambda \in \mathbb{C}.$\\
\indent  Assuming that $u(x,y)$ exists and is sufficiently smooth in $\Omega$, $\mu_j(x,y,\lambda) (j=1,2,3.)$ are the $2\times2$ matrix valued functions
defined by
\begin{equation}
\mu_j(x,y;\lambda)=I+\int_{(x_j,y_j)}^{(x,y)}e^{-i(\lambda^2x+2\lambda^6y)\hat{\sigma}_3}W(\xi,\eta,\lambda),\,\,\,\,(x,y)\in \Omega,j=1,2,3.
\end{equation}
The integral denotes a smooth curve from $(x_j,y_j)$ to $(x,y)$, and
$(x_1, y_1)=(\infty, y ), (x_2, y_2) = (0,L), (x_3, y_3) = (0, 0)$. \\
\indent The fundamental theorem of calculus implies that the functions $\mu_j(x,y;\lambda)(j=1,2,3.)$ satisfy Eq.(2.18).
Since the one-form $W(x,y;\lambda)$ is exact, then $\mu_j(x,y;\lambda)(j=1,2,3.)$ are independent of the path of integration.
We choose the particular contours shown in Fig.2.
\begin{figure}[h]
\centering
\includegraphics[width=1.87in,height=1.87in]{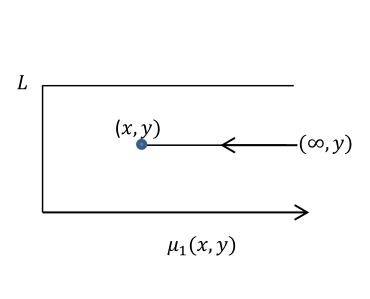}
\includegraphics[width=1.87in,height=1.87in]{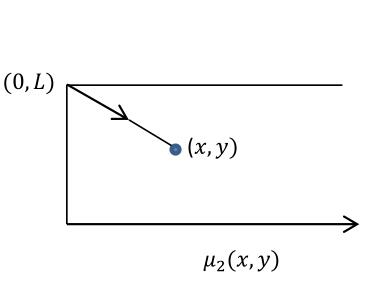}
\includegraphics[width=1.87in,height=1.87in]{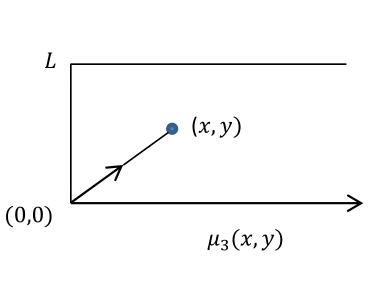}
\caption{The three contours in the $(x,y)$-domaint.}
\label{fig:graph}
\end{figure}
This choice implies the following inequalities on the contours
\begin{equation}
\begin{array}{ll}
(x_1,y_1)\rightarrow (x,y): \xi -x\geq 0,\\
(x_2,y_2)\rightarrow (x,y): \xi -x\leq 0,\eta-y\geq 0,\\
(x_2,y_2)\rightarrow (x,y): \xi -x\leq 0,\eta-y\leq 0.
\end{array}
\end{equation}
The functions $\mu_1$, $\mu_2$ and $\mu_3$ are defined from $\lambda$ in some domain of the complex $\lambda$-plane. Following the idea in Ref.[15], we have
\begin{equation}
\mu_1(x,y;\lambda)=I-\int_{x}^{\infty}e^{-i\lambda^2(x-\xi)\hat{\sigma}_3}(N_1\mu_1)(\xi,y,\lambda)d\xi,\end{equation}
\begin{equation}
\mu_2(x,y ;\lambda)=I+\int_{0}^{x}e^{-i\lambda^2(x-\xi)\hat{\sigma}_3}(N_1\mu_2)(\xi,y,\lambda)d\xi-e^{-i\lambda^2x\hat{\sigma}_3}\int_{y}^{L}e^{-2i\lambda^6(y-\eta)\hat{\sigma}_3}(N_2\mu_2)(0,\eta,\lambda)d\eta,\end{equation}
\begin{equation}
\mu_3(x,y;\lambda)=I+\int_{0}^{x}e^{-i\lambda^2(x-\xi)\hat{\sigma}_3}(N_1\mu_3)(\xi,y,\lambda)d\xi+e^{-i\lambda^2x\hat{\sigma}_3}\int_{0}^{y}e^{-2i\lambda^6(y-\eta)\hat{\sigma}_3}(N_2\mu_3)(0,\eta,\lambda)d\eta.
\end{equation}
We find that the first column of the matrix Eq.(2.26) involves $e^{2i\lambda^2(x-\xi)}$, the first column of the matrix Eq.(2.27) and Eq.(2.28) involve $e^{2i\lambda^2(x-\xi)}$ and $e^{4i\lambda^6(y-\eta)}$.
Using the above inequalities (2.25) implies that the exponential term of $\mu_j(x,y;\lambda)(j=1,2,3.)$ is bounded in the
following regions of the complex $\lambda$-plane,
\begin{equation}
\begin{array}{ll}
\mu_1^{(1)}(x,y;\lambda): \{Im\lambda^2\leq 0\},\\
\mu_2^{(1)}(x,y;\lambda): \{Im\lambda^2\geq 0\}\bigcap\{Im\lambda^6\leq 0\},\\
\mu_3^{(1)}(x,y;\lambda): \{Im\lambda^2\geq 0\}\bigcap\{Im\lambda^6\geq 0\},
\end{array}
\end{equation}
where $\mu_j^{(1)}(x,y;\lambda)(j=1,2,3.)$ denote the the first column of the matrix $\mu_j(x,y;\lambda)(j=1,2,3.)$.
The second column $\mu_j^{(2)}(x,y;\lambda)(j=1,2,3.)$ of the matrix Eq.(2.26) involves the inverse of the above exponential, similar considerations are valid for $\mu_j^{(2)}(x,y;\lambda)(j=1,2,3.)$. We have
\begin{equation}
\begin{array}{ll}
\mu_1^{(2)}(x,y;\lambda): \{Im\lambda^2\geq 0\},\\
\mu_2^{(2)}(x,y;\lambda): \{Im\lambda^2\leq 0\}\bigcap\{Im\lambda^6\geq 0\},\\
\mu_3^{(2)}(x,y;\lambda): \{Im\lambda^2\leq 0\}\bigcap\{Im\lambda^6\leq 0\},
\end{array}
\end{equation}
We define the $D_i(i=1,2,\cdots,6.)$ in the complex $\lambda$-plane by
\begin{equation}
D_i=\{ \lambda \in \mathbb{C}, k \pi+ \frac{(i-1) \pi}{6}< Arg \lambda < k \pi+\frac{i \pi}{6}, i=1,2,\cdots,6.\,\, k=0,1,2,\cdots.  \},
\end{equation} where $Arg \lambda$ denotes the argument of the $\lambda$,
see Fig.3.
\begin{figure}
\centering
\includegraphics[width=2.87in,height=2.87in]{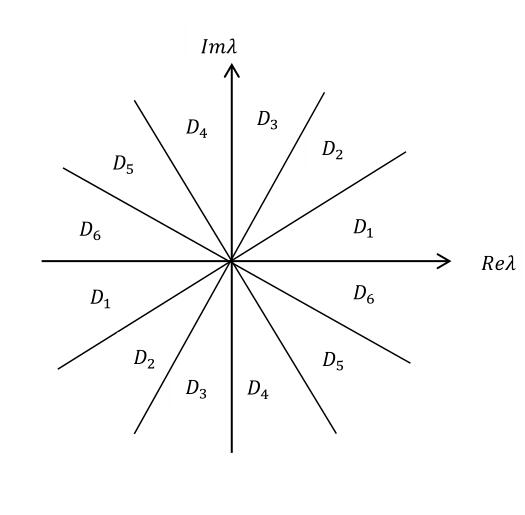}
\caption{The sets $D_j,j=1,2,3,4,5,6$, which decompose the complex $\lambda-$plane}
\label{fig:graph}
\end{figure}
Then, we obtain
\begin{equation}\begin{array}{ll}
\mu_1(x,y;\lambda)=(\mu_{1}^{D_4\cup D_5\cup D_6}(x,y;\lambda),\mu_{1}^{D_1\cup D_2\cup D_3}(x,y;\lambda)),\\
\mu_2(x,y;\lambda)=(\mu_{2}^{D_2}(x,y;\lambda),\mu_{2}^{D_5}(x,y;\lambda)),\\
\mu_3(x,y;\lambda)=(\mu_{3}^{D_1\cup D_3}(x,y;\lambda),\mu_{3}^{D_4\cup D_6}(x,y;\lambda))
\end{array}\end{equation}
where $\mu_{j}^{D_i}(j=1,2,3.i=1,2,\cdots,6.)$ denotes $\mu_j$ which is bounded and analytic for $\lambda \in D_i$.\\
\noindent More specifically,
 \begin{equation}\begin{array}{ll}
\mu_2(0,y;\lambda)=(\mu_{2}^{D_2\cup D_4\cup D_6}(0,y;\lambda),\mu_{2}^{D_1\cup D_3\cup D_5}(0,y;\lambda)),\\
\mu_2(x,L;\lambda)=(\mu_{2}^{D_1\cup D_2\cup D_3}(x,L;\lambda),\mu_{2}^{D_4\cup D_5\cup D_6}(x,L;\lambda)),\\
\mu_2(0,0;\lambda)=(\mu_{2}^{D_2\cup D_4\cup D_6}(0,0;\lambda),\mu_{2}^{D_1\cup D_3\cup D_5}(0,0;\lambda)),\\
\mu_3(0,y;\lambda)=(\mu_{3}^{D_1\cup D_3\cup D_5}(0,y;\lambda),\mu_{3}^{D_2\cup D_4\cup D_6}(0,y;\lambda)),\\
\mu_3(x,0;\lambda)=(\mu_{3}^{D_1\cup D_2\cup D_3}(x,0;\lambda),\mu_{3}^{D_4\cup D_5\cup D_6}(x,0;;\lambda)),\\
\mu_3(0,L;\lambda)=(\mu_{3}^{D_1\cup D_3\cup D_5}(0,L;\lambda),\mu_{3}^{D_2\cup D_4\cup D_6}(0,L;\lambda)).
\end{array}\end{equation}
\textbf{\textbf{2.4.} The Spectral Functions and Their Propositions}\\
\indent The crucial advantage of requiring that the functions $\mu_j(j=1,2,3.)$ solve equation (2.24) is that any
two such functions are related via a matrix that has explicit exponential $(x,y)-$dependence of the form $exp[-i(\lambda^2x+\lambda^6y)\sigma_3] \rho (\lambda)$, where the function $\rho (\lambda)$ can be computed by
evaluating the relevant relation at any convenient point in the domain $\Omega$. In particular, using
that $\mu_3(0,0,\lambda)=I$. In order to  deriving a RHP, we need to compute the jumps across the
boundaries of the $D_j$'s $(j=1,2,\cdots,6.)$. It turns out that the relevant jump matrices can be uniquely defined
in terms of two $2\times2$ matrices valued spectral functions $S_1(\lambda)$ and $S_2(\lambda)$ defined as follows
\begin{equation}\begin{array}{cc}
\mu_1(x,y;\lambda)=\mu_3(x,y;\lambda)e^{-i(\lambda^2x+2\lambda^6y)\hat{\sigma}_3}S_1(\lambda),\\
\mu_2(x,y;\lambda)=\mu_3(x,y;\lambda)e^{-i(\lambda^2x+2\lambda^6y)\hat{\sigma}_3}S_2(\lambda).
\end{array}\end{equation}
Evaluating the first equation of (2.34) at $(x,y)=(0,0)$ and the second equation of (2.34) at $(x,y) = (0,L)$, implies
\begin{equation}
S_1(\lambda)=\mu_1(0,0;\lambda),S_2(\lambda)=(e^{2i\lambda^6L\hat{\sigma}_3}\mu_3(0,L;\lambda))^{-1}
\end{equation}
From Eq.(2.34) and Eq.(2.35), we obtain
\begin{equation}
\mu_2(x,y;\lambda)=\mu_1(x,y;\lambda)e^{-i(\lambda^2x+2\lambda^6y)\hat{\sigma}_3}(S_1(\lambda))^{-1}S_2(\lambda)
\end{equation}
which will lead to the global relation.\\
\indent Hence, the function $S_1(\lambda)$ can be obtained from the evaluations at $x = 0$ of the function
$\mu_1(x, 0, \lambda)$ and $S_2(\lambda)$ can be obtained from the evaluations at $y = L$ of the function $\mu_3(0, y, \lambda)$.  And these
functions about $\mu_j(x,y;\lambda)(j=1,2,3,)$ satisfy the linear integral equations as follows
\begin{equation}
\mu_1(x,0;\lambda)=I-\int_{x}^{\infty}e^{-i\lambda^2(x-\xi)\hat{\sigma}_3}(N_1\mu_1)(\xi,0,\lambda)d\xi,\end{equation}
\begin{equation} \mu_3(x,0;\lambda)=I+\int_{0}^{x}e^{-i\lambda^2(x-\xi)\hat{\sigma}_3}(N_1\mu_3)(\xi,0,\lambda)d\xi,\end{equation}
\begin{equation} \mu_2(0,y;\lambda)=I-\int_{y}^{L}e^{-2i\lambda^6(y-\eta)\hat{\sigma}_3}(N_2\mu_2)(0,\eta,\lambda)d\eta,\end{equation}
\begin{equation} \mu_3(0,y;\lambda)=I+\int_{0}^{y}e^{-2i\lambda^6(y-\eta)\hat{\sigma}_3}(N_2\mu_3)(0,\eta,\lambda)d\eta.
\end{equation}
Let $u_0(x)=u(x,0)$, $g_0(y)=u(0,y)$, $g_1(y)=u_x(0,y)$ and $g_2(y)=u_{xx}(0,y)$ be the initial and boundary values of $u(x,y)$, then
\begin{equation}
N_1(x,0;\lambda)=\left(
                   \begin{array}{cc}
                     -iu_0(x) & \lambda u_0(x)e^{-i\int_{0}^{x}u_0(\xi)d\xi} \\
                     2\lambda e^{-i\int_{0}^{x}u_0(\xi)d\xi}  &  iu_0(x) \\
                   \end{array}
                 \right),
\end{equation}
\begin{equation}
N_2(0,y;\lambda)=\left(
                   \begin{array}{cc}
                     N_2^{11}(0,y;\lambda) & N_2^{12}(0,y;\lambda) \\
                     N_2^{21}(0,y;\lambda) &-N_2^{11}(0,y;\lambda) \\
                   \end{array}
                 \right),
\end{equation}
where\\
$$
\begin{array}{ll}
N_2^{11}(0,y;\lambda)=-2i\lambda^4g_0(y)+\lambda^2(g_1(y)-i(g_0(y))^2)+\frac{i}{2}g_2(y)+\frac{3}{2}g_0(y)g_1(y)-\frac{i}{2}(g_0(y))^3,\\
N_2^{12}(0,y;\lambda)=(2\lambda^5g_0(y)+\lambda^3(ig_1(y)+g_0^2(y))+\lambda(-\frac{1}{2}g_2(y)+\frac{3}{2}ig_0(y)g_1(y)+\frac{1}{2}(g_0(y))^3) e^{-2i\int_{0}^{y}\Delta_2(0,\eta) d\eta},\\
N_2^{21}(0,y;\lambda)=(4\lambda^5+2\lambda^3g_0(y)+\lambda(ig_1(y)+g_0^2(y)) e^{-2i\int_{0}^{y}\Delta_2(0,\eta) d\eta},\\
\Delta _2(0,y)=-\frac{1}{4}g_2(y)+\frac{3}{4}ig_0(y)g_1(y)+\frac{1}{4}g_0^3(y).
\end{array}
$$\\
These expressions for $N_1(x,0,\lambda), N_2 (0,y,\lambda) $ contain only $u_0(x),g_0 (y), g_1 (y)$ and
$g_2(y)$, respectively. Therefore, the integral equation (2.37) determining $S_1(\lambda)$ is defined
in terms of the initial data $u_0(x)$, the integral equation (2.40) determining $S_2(\lambda)$ is defined in
terms of the initial data $g_0(y), g_1(y)$ and $g_2(y)$.\\
\indent Following the Section 2.4. of the Ref.[9], let us show the function $\mu(x,y,\lambda)$ satisfy the symmetry relations. The analytic properties of $2\times2$ matrices $\mu_j(x,y;\lambda)(j=1,2,3.)$ that come from Eq.(2.26), Eq.(2.27) and Eq.(2.28) are collected in the following proposition.
Setting
\begin{center}
$\mu_j(x,y;\lambda)=(\mu_j^{(1)}(x,y;\lambda),\mu_j^{(2)}(x,y;\lambda))=\left(
                                                                          \begin{array}{cc}
                                                                            \mu_j^{11} & \mu_j^{12} \\
                                                                            \mu_j^{21} & \mu_j^{22} \\
                                                                          \end{array}
                                                                        \right),j=1,2,3.
$
\end{center}
\textbf{Proposition 2.4.1.} The matrices $\mu_j(x,y;\lambda)=(\mu_j^{(1)}(x,y;\lambda),\mu_j^{(2)}(x,y;\lambda)) (j=1,2,3.)$ have the following properties
\begin{description}
  \item [(1)]$det\mu_1(x,y;\lambda)=det\mu_2(x,y;\lambda)=det\mu_3(x,y;\lambda)=1$;
  \item [(2)]$\mu_1^{(1)}(x,y;\lambda)$ is analytic, and $\lim \limits_ {\lambda\rightarrow\infty}\mu_1^{(1)}(x,y;\lambda)=(1,0)^T,\,\,\,\lambda\in\{Im\lambda^2 < 0\}$;
  \item [(3)]$\mu_1^{(2)}(x,y;\lambda)$ is analytic, and $\lim \limits_ {\lambda\rightarrow\infty}\mu_1^{(2)}(x,y;\lambda)=(0,1)^T,\,\,\,\lambda\in\{Im\lambda^2 > 0\}$;
  \item [(4)]$\mu_2^{(1)}(x,y;\lambda)$ is analytic, and $\lim \limits_ {\lambda\rightarrow\infty}\mu_2^{(1)}(x,y;\lambda)=(1,0)^T,\,\,\,\lambda\in\{Im\lambda^2 > 0\}\bigcap\{Im\lambda^6 < 0\}$;
  \item [(5)]$\mu_2^{(2)}(x,y;\lambda)$ is analytic, and $\lim \limits_ {\lambda\rightarrow\infty}\mu_2^{(2)}(x,y;\lambda)=(0,1)^T,\,\,\,\lambda\in\{Im\lambda^2 < 0\}\bigcap\{Im\lambda^6 > 0\}$;
  \item [(6)]$\mu_3^{(1)}(x,y;\lambda)$ is analytic, and $\lim \limits_ {\lambda\rightarrow\infty}\mu_3^{(1)}(x,y;\lambda)=(1,0)^T,\,\,\,\lambda\in\{Im\lambda^2 > 0\}\bigcap\{Im\lambda^6 > 0\}$;
  \item [(7)]$\mu_3^{(2)}(x,y;\lambda)$ is analytic, and $\lim \limits_ {\lambda\rightarrow\infty}\mu_3^{(2)}(x,y;\lambda)=(0,1)^T,\,\,\,\lambda\in\{Im\lambda^2 < 0\}\bigcap\{Im\lambda^6 < 0\}$.
  \end{description}
\textbf{Proposition 2.4.2.(Symmetries)} The matrices $\mu_j(x,y;\lambda)=\left(
                                                                          \begin{array}{cc}
                                                                            \mu_j^{11}(x,y;\lambda) & \mu_j^{12}(x,y;\lambda) \\
                                                                            \mu_j^{21}(x,y;\lambda) & \mu_j^{22}(x,y;\lambda) \\
                                                                          \end{array}
                                                                        \right)(j=1,2,3.)
$ have the following properties

\begin{description}
  \item [(1)]$\mu_j^{11}(x,y;\lambda)=\overline{\mu_j^{22}(x,y;\bar{\lambda})}$, $\mu_j^{12}(x,y;\lambda)=\overline{\mu_j^{21}(x,y;\bar{\lambda})}$;
  \item [(2)]$\mu_j^{11}(x,y;-\lambda)=\mu_j^{11}(x,y;\lambda)$, $\mu_j^{12}(x,y;-\lambda)=-\mu_j^{12}(x,y;\lambda)$,\\$\mu_j^{21}(x,y;-\lambda)=-\mu_j^{21}(x,y;\lambda)$, $\mu_j^{22}(x,y;-\lambda)=\mu_j^{22}(x,y;\lambda)$.

\end{description}
\textbf{Proposition 2.4.3.} The spectral function $S_1(\lambda)$ and $S_2(\lambda)$ are defined in Eq.(2.34) and Eq.(2.35) imply that
\begin{equation}
\begin{array}{cc}
S_1(\lambda)=I-\int_{0}^{\infty}e^{-i\lambda^2(x-\xi)\hat{\sigma}_3}(N_1\mu_1)(\xi,0;\lambda)d\xi,\\
S_2^{-1}(\lambda)=I+\int_{0}^{L}e^{2i\lambda^6\eta\hat{\sigma}_3}(N_2\mu_3)(0,\eta;\lambda)d\eta.
\end{array}
\end{equation}
\indent If $\psi(x,y,\lambda)$ satisfies Eq.(2.3), it follows that $det\psi$ is independent
of $x$ and $y$. Hence, since $detD(x,y)=1$, the determinant of
the function $\mu(x,y,\lambda)$ corresponding to $\psi$ according to (2.17) is also
independent of $x$ and $y$. In particular, for $\mu_ j, j = 1 , 2 , 3.$ evaluation
of $det\mu_ j$ at $(x,y)$ shows that
$\mu_ j=I+o(\frac{1}{\lambda})( j = 1 , 2 , 3.)$ and $det\mu_j=1( j = 1 , 2 , 3.)$.
According to \textbf{Proposition 2.4.2}, we can construct the following matrix spectral functions $S_1(\lambda)$ and $S_2(\lambda)$,
\begin{equation}
S_1(\lambda)=\left(
             \begin{array}{cc}
               \overline{a(\bar{\lambda})} & b(\lambda)\\
               \overline{b(\bar{\lambda})}& a(\lambda)\\
             \end{array}
           \right), \,\,\,\,
S_2(\lambda)=\left(
             \begin{array}{cc}
               \overline{A(\bar{\lambda})} & B(\lambda)\\
               \overline{B(\bar{\lambda})}& A(\lambda)\\
             \end{array}
           \right)
\end{equation}
By use of  Eq.(2.35), we can obtain the proposition about the matrix spectral functions  (2.44).\\
\textbf{Proposition 2.4.4.}The spectral function $S_1(\lambda)$ and $S_2(\lambda)$ are constructed in Eq.(2.44) have the following properties
\begin{itemize}
  \item[(1)]
$\left(
   \begin{array}{c} b(\lambda) \\ a(\lambda) \\ \end{array} \right)=\mu_1^{(2)}(0,0;\lambda)=\left(
   \begin{array}{c}
    \mu_1^{12}(0,0;\lambda) \\
    \mu_1^{22}(0,0;\lambda) \\
  \end{array}
 \right). \\$
 \item[(2)]
$\left(
  \begin{array}{c} -e^{-4i\lambda^6L}B(\lambda) \\ \overline{A(\bar{\lambda})} \\ \end{array} \right)=\mu_3^{(2)}(0,L;\lambda)=\left(
  \begin{array}{c}
    \mu_3^{12}(0,L;\lambda) \\
    \mu_3^{22}(0,L;\lambda) \\
  \end{array}
\right).$
 \item [(3)]
$\partial_x\mu_1^{(2)}(x,0;\lambda)-2i\lambda^2 \tilde{\sigma} \mu_1^{(2)}(x,0;\lambda)=N_1(x,0;\lambda)\mu_1^{(2)}(x,0;\lambda),\,\,\,\lambda\in D_1\bigcup D_2\bigcup D_3,0<x<\infty,\\
  \partial_y\mu_3^{(2)}(0,y;\lambda)+4i\lambda^6\tilde{\sigma}\mu_3^{(2)}(0,y;\lambda)=N_2(0,y;\lambda)\mu_3^{(2)}(0,y;\lambda),\,\,\,\lambda\in D_2\bigcup D_4\bigcup D_6,0<y<L,\\
where \tilde{\sigma}=\left(
                       \begin{array}{cc}
                         1 & 0 \\
                         0 & 0 \\
                       \end{array}
                     \right).$
 \item [(4)] $a(\lambda)$,$b(\lambda)$,$A(\lambda)$ and $B(\lambda)$ obey the symmetries $a(-\lambda)=a(\lambda),\,\,\,b(-\lambda)=-b(\lambda),
  A(-\lambda)=A(\lambda),\,\,\,B(-\lambda)=-B(\lambda).$
 \item[(5)]$ det S_1(\lambda)=det S_2(\lambda)=1.$
 \item [(6)]$a(\lambda)=1+\sum_{j=1}^m \frac{a_j(\lambda)}{\lambda^m}+O(\frac{1}{\lambda^{m+1}}),\,\,\,b(\lambda)=\sum_{j=1}^m \frac{b_j(\lambda)}{\lambda^m}+O(\frac{1}{\lambda^{m+1}})$, uniformly as $\lambda\rightarrow\infty$ with $Im\lambda^2> 0.$
$A(\lambda)=1+\sum_{j=1}^m \frac{A_j(\lambda)}{\lambda^m}+O(\frac{1}{\lambda^{m+1}}),\,\,\,B(\lambda)=\sum_{j=1}^m \frac{B_j(\lambda)}{\lambda^m}+O(\frac{1}{\lambda^{m+1}})$, uniformly as $\lambda\rightarrow\infty$ with $Im\lambda^6> 0.$
 \end{itemize}
\textbf{2.5. The Global Relation}\\
\indent Let $F(y,\lambda)$ and $G(y,\lambda)$ denote the eigenfunction obtained from the evaluation of $\mu_1(x,y,\lambda)$ and $\mu_3(x,y,\lambda)$ at $x = 0$, respectively
\begin{equation}
F(y,\lambda)=\mu_1(0,y,\lambda) ,\,\,\,\,G(y,\lambda)=\mu_3(0,y,\lambda),\,\,\,\,\lambda\in \mathbb{C}.
\end{equation}
The eigenfunctions $F(y,\lambda)$ and $G(y,\lambda)$ satisfy the differential Eq.(2.23) thus, they are simply related
\begin{equation}
G(y,\lambda)e^{-2i\lambda^6y\hat{\sigma}_3}S_1(\lambda)=F(y,\lambda),\,\,\,\,0<y<L,\lambda\in \mathbb{C}.
\end{equation}
Evaluating this equation at $y=L$, we obtain the global relation
\begin{equation}
S_2^{-1}(\lambda)S_1(\lambda)=e^{2i\lambda^6L\hat{\sigma}_3}F(L,\lambda),\,\,\,\,\lambda\in \mathbb{C}.
\end{equation}
\textbf{Remark 2.5.1.} For the NLS, there exists one unknown boundary value. Thus, one might expect
that since the global relation (2.24) provides one equation connecting this unknown function
with the given initial and boundary conditions, in this case, it is possible, by utilizing the global
relation, to characterize the unknown boundary value. However, for the mKdV, there exist two
unknown boundary values. In this case, the solution of the associated linear problem suggests
that it is impossible to solve this problem, unless one uses the transformation that leaves $\lambda^6$
invariant in order to obtain an additional equation from the global relation.\\
\textbf{Remark 2.5.2.} It is important to note that the rigorous result mentioned earlier does not require
the knowledge of the explicit form of $F(y,\lambda)$, it only requires the existence of a function $F(y,\lambda)$ with
specific analyticity properties. This suggests that the most efficient approach of characterizing
${A(\lambda),B(\lambda)}$ is to actually eliminate $F(y,\lambda)$. This is precisely the philosophy used in the linear limit
for the derivation of the generalized Dirichlet to Neumann map and it is also the philosophy
used in [5,8].\\
\textbf{2.6. The Basic RHP}\\
\indent Following, according to the paper [16], we can get that the RHP of the cSTO equation. Equation (2.34) relating the various analytic eigenfunctions, can be rewritten
in a form that determines the jump conditions of a $2\times2$ RHP, with unitary
jump matrices on the real and imaginary axis. This involves tedious but straight forward
algebraic manipulations. The final form is \\
\begin{equation}
M_+(x,y,\lambda)=M_-(x,y,\lambda)J(x,y,\lambda),\,\,\,\,\lambda^6\in\mathbb{R}
\end{equation}
where $M_{+}(x,y,\lambda)$ and  $M_{-}(x,y,\lambda)$ have two forms of expression, respectivly
\begin{equation}
\begin{array}{ll}
M^{(1)}_+(x,y,\lambda)=(\frac{\mu_1^{D_4\cup D_5\cup D_6}(x,y,\lambda)}{\alpha(\lambda)},\mu_2^{D_5}(x,y,\lambda)),\,\,\,\,\,\,\,\,\,\,\,\,\,\,\,\lambda\in D_5,\\
M^{(1)}_{-}(x,y,\lambda)=(\frac{\mu_1^{D_4\cup D_5\cup D_6}(x,y,\lambda)}{\overline{a(\bar{\lambda})}},\mu_3^{D_4\cup D_6}(x,y,\lambda)), \lambda\in D_4\cup D_6,\\
M^{(2)}_{+}(x,y,\lambda)=(\mu_3^{D_1\cup D_3}(x,y,\lambda),\frac{\mu_1^{D_1\cup D_2\cup D_3}(x,y,\lambda)}{a(\lambda)}),\,\,\,\,\,\,\,\,\, \lambda\in D_1\cup D_3,\\
M^{(2)}_{-}(x,y,\lambda)=(\mu_2^{D_2}(x,y,\lambda),\frac{\mu_1^{D_1\cup D_2\cup D_3}(x,y,\lambda)}{\overline{\alpha(\bar{\lambda})}}),\,\,\,\,\,\,\,\,\,\,\,\,\,\,\, \lambda\in D_2.
\end{array}
\end{equation}
Equation (2.46) imply that
$$\begin{array}{cc}
detM(x,y;\lambda)=1,\\
M(x,y;\lambda)=I+O(\frac{1}{\lambda}),\,\,\lambda\rightarrow\infty.
\end{array}$$
Setting
$
\theta(\lambda)=\lambda^2x+2\lambda^6y,
\alpha(\lambda)=\overline{a(\bar{\lambda})}A(\lambda)-\overline{b(\bar{\lambda})}B(\lambda),
 \beta(\lambda)=a(\lambda)B(\lambda)-b(\lambda)A(\lambda)
$, we have the following theorem\\
\textbf{Theorem2.6.1.} Let $u(x,y;\lambda)$ is a smooth function, $\mu_1(x,y,\lambda),\mu_2(x,y,\lambda),\mu_3(x,y,\lambda)$ are defined by Eq.(2.26), Eq.(2.27) and Eq.(2.28), $M(x,y;\lambda)$ be defined by Eq.(2.46), then $M(x,y;\lambda)$
satisfies the jump condition
\begin{equation}
M_{+}(x,y,\lambda)=M_{-}(x,y,\lambda)J(x,y,\lambda),\,\,\,\,\lambda^6\in\mathbb{R},
\end{equation}
where
\begin{equation}
J(x,y,\lambda)=\left\{
                 \begin{array}{ll}
                   J_1(x,y,\lambda), & \hbox{$Arg \lambda=k\pi \,\,\,or\,\,\, k\pi+\frac{\pi}{2} $;} \\
                   J_2(x,y,\lambda), & \hbox{$Arg \lambda=k\pi+\frac{\pi}{6} \,\,\,or\,\,\, k\pi+\frac{\pi}{3}$;} \\
                   J_3(x,y,\lambda), & \hbox{$Arg \lambda=k\pi+\frac{2\pi}{3} \,\,\,or\,\,\, k\pi+\frac{5\pi}{6}$.} \\
                 \end{array}
               \right.
\end{equation}
 $$\begin{array}{cc}
J_1(x,y,\lambda)=\left(
                   \begin{array}{cc}
                    1  & \frac{b(\lambda)}{\overline{a(\bar{\lambda})}}e^{-2i\theta(\lambda)} \\
                    -\frac{\overline{b(\bar{\lambda})}}{a(\lambda)}e^{2i\theta(\lambda)} & \frac{1}{a(\lambda)\overline{a(\bar{\lambda})}} \\
                   \end{array}
                 \right),\\
J_2(x,y,\lambda)=\left(
                   \begin{array}{cc}
                     \frac{a(\lambda)}{\overline{\alpha(\bar{\lambda})}} & 0 \\
                    -e^{2i\theta(\lambda)}\overline{B(\bar{\lambda})} & \frac{\overline{\alpha(\bar{\lambda})}}{a(\lambda)} \\
                   \end{array}
                 \right),\\
J_3(x,y,\lambda)=\left(
                   \begin{array}{cc}
                     \frac{\overline{a(\overline{\lambda})}}{\alpha(\lambda)} & B(\lambda)e^{-2i\theta(\lambda)} \\
                    0 & \frac{\alpha(\lambda)}{\overline{a(\bar{\lambda})}} \\
                   \end{array}
                 \right).\\
\end{array}$$
The contour for this RHP is depicted in Fig.4.
\begin{figure}
\centering
\includegraphics[width=3.87in,height=3.87in]{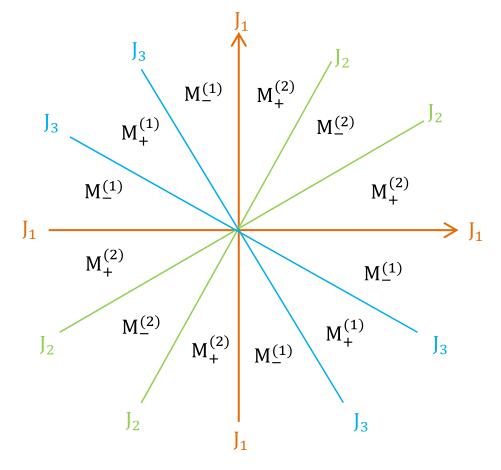}
\caption{The contour for the RHP on the complex $\lambda-$ plane.}
\label{fig:graph}
\end{figure}\\
\textbf{Proof.} Using the idea in ref.[17], in order to derive the jump condition Eq.(2.47) we write Eq.(2.34) and Eq.(2.36) in the following form
\begin{equation}\left\{
  \begin{array}{ll}
   \overline{a(\bar{\lambda})}\mu_3^{D_1\cup D_3}+\overline{b(\bar{\lambda})}e^{2i\theta(\lambda)}\mu_3^{D_4\cup D_6}=\mu_1^{D_4\cup D_5\cup D_6}  \\
   b(\lambda)e^{-2i\theta(\lambda)}\mu_3^{D_1\cup D_3}+a(\lambda)\mu_3^{D_4\cup D_6}=\mu_1^{D_1\cup D_2\cup D_3},
  \end{array}
\right.\end{equation}
\begin{equation}\left\{
  \begin{array}{ll}
   \overline{A(\bar{\lambda})}\mu_3^{D_1\cup D_3}+\overline{B(\bar{\lambda})}e^{2i\theta(\lambda)}\mu_3^{D_4\cup D_6}=\mu_2^{D_2}  \\
   B(\lambda)e^{-2i\theta(\lambda)}\mu_3^{D_1\cup D_3}+A(\lambda)\mu_3^{D_4\cup D_6}=\mu_2^{D_5},
  \end{array}
\right.\end{equation}
\begin{equation}\left\{
  \begin{array}{ll}
   \overline{\alpha(\bar{\lambda})}\mu_1^{D_4\cup D_5\cup D_6}+\overline{\beta(\bar{\lambda})}e^{2i\theta(\lambda)}\mu_1^{D_1\cup D_2\cup D_3}=\mu_2^{D_2}  \\
   \beta(\lambda)e^{-2i\theta(\lambda)}\mu_1^{D_4\cup D_5\cup D_6}+\alpha(\lambda)\mu_1^{D_1\cup D_2 \cup D_3}=\mu_2^{D_5}.
  \end{array}
\right.\end{equation}
By direct calculation, we can  derive that the jump matrices $J_i(x,y;\lambda)(i=1,2,3.)$ satisfy the jump condition Eq.(2.47).\\
\indent Further more, we can obtain the jump condition between $M_+^{(1)}$ and $M_-^{(2)}$
\begin{equation}
M_+^{(1)}=M_-^{(2)}J_4
\end{equation}
where $J_4=J_2J_1^{-1}J_3$.\\
\textbf{2.7. The Residue Conditions.}\\
\indent The matrix $M(x,y;\lambda)$ of this RHP is a sectionally meromorphic function
of $\lambda$ in $\mathbb{C} \setminus \{\lambda^6 \in \mathbb{R}\}$. The possible poles of $M(x,y;\lambda)$ are generated by the zeros of $a(\lambda)$, $\alpha(\lambda)$ and by the complex
conjugates of these zeros. Since $a(\lambda)$, $\alpha(\lambda)$ are even functions, this means each zero $\lambda_j$ of $a(\lambda)$ is accompanied by another zero at $-\lambda_j$.
Similarly, each zero $\lambda_j$ of $\alpha(\lambda)$ is accompanied by a zero at $-\lambda_j$. In particular, both $a(\lambda)$ and
$\alpha(\lambda)$ have even number of zeros.\\

\textbf{Hypothesis 2.7.1. }  We assume that
\begin{itemize}
 \item $a(\lambda)$ has $2\wedge$ simple zeros($2\wedge=2\wedge_1+2\wedge_2$), such that $\zeta_j$( $j=1,2,\cdots,2\wedge_1$) lie in $D_1\bigcup D_3$, and $\bar{\zeta}_j$($j=2\wedge_1+1,2\wedge_1+2,\cdots,2\wedge$) lie in $D_4\bigcup D_6$.
  \item $\alpha(\lambda)$ has $2\vee$ simple zeros($2\vee=2\vee_1+2\vee_2$), such that $\varepsilon_j$($j=1,2,\cdots,2\vee_1$), lie in $D_5$, and $\bar{\varepsilon}_j$, $j=2\vee_1+1,2\vee_1+2,\cdots,2\vee$, lie in $D_2$.
  \item None of the zeros of $\alpha(\lambda)$ coincides with any of the zeros of $a(\lambda)$.
\end{itemize}
The residues of the function $M(x,y;\lambda)$ at the corresponding poles can be computed using Eq.(2.34) and Eq.(2.36). Using the notation $[M_{\pm}^{(i)}(x,y;\lambda)]_1(i=1,2.)$ for the first column and $[M_{\pm}^{(i)}(x,y;\lambda)]_2(i=1,2.)$ for the second column of the solution $M_{\pm}^{(i)}(x,y;\lambda)(i=1,2.)$ of the RHP, and we write $\dot{\alpha}(\lambda)=\frac{d\alpha}{d\lambda}$, then we get the following proposition.\\
\\
\textbf{Proposition 2.7.2.}
\begin{description}
  \item[(1)] Res $\{[M_+^{(2)}(x,y;\lambda)]_{2} ,  \zeta_j \}$=$\frac{b(\zeta_j)}{\dot{a}(\zeta_j)}e^{-2i\theta(\zeta_j)}[M_+^{(2)}(x,y;\zeta_j)]_{1}$, $j=1,2,\cdots,2\wedge_1$.
  \item[(2)] Res $\{[M_-^{(1)}(x,y;\lambda)]_{1} ,  \bar{\zeta_j} \}$=$\frac{\overline{b(\zeta_j)}}{\overline{\dot{a}(\zeta_j)}}e^{2i\theta(\bar{\zeta_j})}[M_-^{(1)}(x,y;\bar{\zeta_j})]_{2}$, , $j=2\wedge_1+1,2\wedge_1+2,\cdots,2\wedge$.
\item[(3)] Res $\{[M_+^{(1)}(x,y;\lambda)]_{1} ,  \varepsilon_j \}$=$-\frac{\overline{\beta(\bar{\varepsilon}_j)}}{\dot{\alpha}(\varepsilon_j)}e^{2i\theta(\varepsilon_j)}[M_+^{(1)}(x,y;\varepsilon_j)]_{2}$, $j=1,2,\cdots,2\vee_1$.
\item[(4)] Res $\{[M_-^{(2)}(x,y;\lambda)]_{2} ,  \bar{\varepsilon}_j \}$=$-\frac{\beta(\bar{\varepsilon}_j)}{\overline{\dot{\alpha}(\varepsilon_j)}}e^{-2i\theta(\bar{\varepsilon}_j)}[M_-^{(2)}(x,y;\bar{\varepsilon}_j)]_{1}$, $j=2\vee_1+1,2\vee_1+2,\cdots,2\vee$.
\end{description}
\textbf{Proof}.\\
\indent According to the idea in ref.[17], we only need to prove \textbf{(1)}, and another three relations also have similar proof.\\
\indent Consider $M_+^{(2)}(x,y;\lambda)]=(\mu_3^{D_1\cup D_3},\frac{\mu_1^{D_1\cup D_2\cup D_3}}{a(\lambda)})$, the simple zeros $\zeta_j$ ($j=1,2,\cdots,2\wedge_1$) of $a(\lambda)$ are the simple poles of $\frac{\mu_1^{D_1\cup D_2\cup D_3}}{a(\lambda)}$. Then we have\\
\begin{equation}\begin{array}{cc}
Res\{\frac{\mu_1^{D_1\cup D_2\cup D_3}}{a(\lambda)},\zeta_j\}=\lim_{\lambda\rightarrow\zeta_j}(\lambda-\zeta_j)\frac{\mu_1^{D_1\cup D_2\cup D_3}(x,y;\lambda)}{a(\lambda)}=\frac{\mu_1^{D_1\cup D_2\cup D_3}(x,y;\zeta_j)}{\dot{a}(\zeta_j)},\,\,\,\, j=1,2,\cdots,2\wedge_1.\end{array}\end{equation}
Taking $\lambda=\zeta_j$ into the second equation of Eq.(2.52) we obtain
\begin{equation}\begin{array}{cc}
 \mu_1^{D_1\cup D_2\cup D_3}(x,y;\zeta_j)=[e^{-2i\theta(\lambda)}b(\lambda)\mu_3^{D_1\cup D_3}(x,y;\lambda)+a(\lambda)\mu_3^{D_4\cup D_6}(x,y;\lambda)]_{\lambda=\zeta_j} \\
= e^{-2i\theta(\zeta_j)}b(\zeta_j)\mu_3^{D_1\cup D_3}(x,y;\zeta_j),\,\,\,\, j=1,2,\cdots,2\wedge_1.
\end{array}\end{equation}
Furthermore,
\begin{equation}
\begin{array}{ll}
 Res \{[M_+^{(2)}(x,y;\lambda)]_{2} ,  \zeta_j \}=Res\{\frac{\mu_1^{D_1\cup D_2\cup D_3}}{a(\lambda)},\zeta_j\}=\frac{e^{-2i\theta(\zeta_j)}b(\zeta_j)\mu_3^{D_1\cup D_3}(x,y;\zeta_j)}{\dot{a}(\zeta_j)}\\
=\frac{b(\zeta_j)}{\dot{a}(\zeta_j)}e^{-2i\theta(\zeta_j)}[M_+^{(2)}(x,y;\zeta_j)]_{1},\,\,\,\, j=1,2,\cdots,2\wedge_1.
\end{array}
\end{equation}
It is equivalent to \textbf{Proposition 2.7.2(1)}.\\
\section {Inverse Spectral Mappings}
\indent~~~~ In this section we turn to the description of the inverse spectral mappings. They will be given in terms of the solutions of the associated RHP, the jump matrices
of which are constructed from the corresponding spectral functions. The guideline for the choice of a particular configuration of these ※boundary§ RHP is that they must be conveniently related to the ※bulk§ RHP, to be constructed for the solution of the CSTO equation in the whole domain. Following this idea, we rewrite the jump condition Eq.(2.50)

\begin{equation}
M_+(x,y;\lambda)-M_{-}(x,y;\lambda)=M_-\tilde{J}(x,y;\lambda),
\end{equation}
where $\tilde{J}(x,y;\lambda)=J(x,y;\lambda)-I$. The asymptotic conditions of Eq.(2.26) Eq.(2.27) Eq.(2.28) and the \textbf{Proposition 2.4.1} imply
\begin{equation}
M(x,y;\lambda)=I+\frac{\overline{M}(x,y;\lambda)}{\lambda}+O(\frac{1}{\lambda}),\,\,\,\,\lambda\rightarrow\infty,\,\,\,\lambda\in \mathbb{C} \setminus \Gamma,
\end{equation}
where $\Gamma=\{Im\lambda^6=0\}$.
Eq.(3.1) and the condition Eq.(3.2) yield the following integral representation
for the function $M(x,y;\lambda)$
\begin{equation}
M(x,y;\lambda)=I+\frac{1}{2\pi i}\int_{\Gamma}\frac{M_{+}(x,y;\lambda')\tilde{J}(x,y;\lambda')}{\lambda-\lambda'}d\lambda',\,\,\,\lambda\in \mathbb{C} \setminus \Gamma,
\end{equation}
then
\begin{equation}
\overline{M}(x,y;\lambda)=-\frac{1}{2\pi i}\int_{\Gamma}M_{+}(x,y;\lambda')\tilde{J}(x,y;\lambda')d\lambda'.
\end{equation}
Using Eq.(3.2) and the ODE of the Lax pair (2.5), we find
\begin{equation} -[\sigma_3,\overline{M}(x,y;\lambda)]=(u_x(x,y)-iu_y(x,y))\sigma_1,\end{equation}
\begin{equation} u_x(x,y)-iu_y(x,y)=2(\overline{M(x,y;\lambda)})_{21}=2\lim_{\lambda\rightarrow\infty}(\lambda M(x,y;\lambda))_{21},\end{equation}
where $\sigma_3$ denote the usual Pauli matrix and $\sigma_1=\left(
                                                               \begin{array}{cc}
                                                                 0 & 1 \\
                                                                 1 & 0 \\
                                                               \end{array}
                                                             \right)
$.\\
\indent The inverse problem involves reconstructing the potential $u(x,y)$ from the spectral functions
$\mu_j$, $j=1,2,3.$ That means we will reconstruct the potential $u(x,y)$. We
show in \textbf{Section 2.2} that
$$\Psi_1^{(o)}=\left(\begin{array}{cc}
                                                                        0&  -\frac{i}{2}uD_{22} \\
                                                                        iD_{11} &0  \\
                                                                      \end{array}
                                                                    \right),$$
when
$$\Psi(x,y;\lambda)=D+\sum_{j=1}^5\frac{\Psi_j(x,y;\lambda)}{\lambda^j}+O(\frac{1}{\lambda^6}), \,\,(\lambda\rightarrow\infty,\,\,\,j=1,2,\cdots,5.)$$
is a solution of Eq.(2.5). This implies that

\begin{equation}
u(x,y)=2im(x,y)e^{2i\int_{(0,0)}^{x,y}\Delta},
\end{equation}
where
$$\mu(x,y;\lambda)=I+\sum_{j=1}^5\frac{m^{(j)}(x,y;\lambda)}{\lambda^j}+O(\frac{1}{\lambda^6}),\,\,(\lambda\rightarrow\infty,\,\,\,j=1,2,\cdots,5.)$$
is the corresponding solution of Eq.(2.18) related to $\Psi(x,y;\lambda)$ via Eq.(2.18), and we write $m(x,y)$ for $m^{(1)}_{12}(x,y)$. From Eq.(3.7) and its complex conjugate $\bar{u}(x,y)$, we obtain
\begin{equation} u\bar{u}=4|m|^2,
u_x\bar{u}-u\bar{u}_x=4(m_x\bar{m}-\bar{m}_xm)+16i|m|^4.\end{equation}
Then we can solve the inverse problem as follows
\begin{description}
  \item[(1)] Use any one of the three spectral functions $\mu_j(j=1,2,3.)$ to
compute $m(x,y)$ according to $$m(x,y)=\lim_{\lambda\rightarrow\infty}(\lambda\mu_j(x,y;\lambda))_{12}.$$
  \item[(2)] Determine $\Delta(x,y)$ from Eq.(2.16).
  \item[(3)] Finally, $u(x,y)$ is given by Eq.(3.7).
\end{description}
\section{The Spectral Functions  and the principal RHP}
\textbf{4.1.The spectral functions $\{a(\lambda),b(\lambda)\}$ }\\
\indent The analysis of \textbf{Section 2} motivates the following definitions for
the spectral functions.\\
\textbf{Definition 4.1.1.} (The spectral functions $a(\lambda)$ and $b(\lambda)$ Given the smooth function $u_{0}(x)=u(x,0)$,
we define the map
\begin{equation} \mathbb{S}: \{u_0(x)\}\rightarrow \{a(\lambda),b(\lambda) \}\end{equation}
by
\begin{equation}
\left(
  \begin{array}{c}
    b(\lambda) \\
    a(\lambda) \\
  \end{array}
\right)=\mu_1^{(2)}(0,\lambda)=\left(
  \begin{array}{c}
    \mu_1^{12}(0,\lambda) \\
    \mu_1^{22}(0,\lambda) \\
  \end{array}
\right),\,\,\,  Im\lambda^2\geq 0.
\end{equation}
where $\mu_1(0,\lambda)$ is the unique solution of the Volterra linear integral equation
$$\mu_1(x,\lambda)=I-\int_{x}^{\infty}e^{-i\lambda^2(x-\xi)\hat{\sigma}_3}(N_1\mu_1)(\xi,0;\lambda)d\xi$$
and $N_1(x, 0;\lambda)$ is given in terms of $u_0(x)$ by Eq.(2.41). The spectral functions $a(\lambda)$ and $b(\lambda)$ have the
same properties with \textbf{Proposition 2.4.4}.\\
\textbf{Definition 4.1.2.}
We define $\mathbb{Q}$ is the inverse map of $\mathbb{S}$,
\begin{equation}
\mathbb{Q}: \{a(\lambda),b(\lambda) \}\rightarrow \{u_0(x)\}
\end{equation} where
$
u_0(x)=2im(x)e^{2i\int_{0}^{x}\Delta_1(\xi)d\xi},
m(x)=\lim_{\lambda\rightarrow\infty}(\lambda M^{(x)}(x,\lambda))_{12}.\\
$
\textbf{Theorem 4.1.3}
\begin{itemize}
  \item $M^{(x)}(x,\lambda)$ is the unique solution of the following RHP\\
$M^{(x)}(x,\lambda)=\left\{
                             \begin{array}{ll}
                              M_{-}^{(x)} (x,\lambda), & Im \lambda^2 <0 \\
                              M_{+}^{(x)} (x,\lambda), & Im \lambda^2 >0
                             \end{array}
                           \right.$ is a sectionally meromorphic function, then $M^{(x)}(x,\lambda)$ satisfied the jump condition
 \begin{equation}M_{+}^{(x)} (x,\lambda)=M_{-}^{(x)} (x,\lambda)J^{(x)} (x,\lambda),\,\,\,\lambda^2\in \mathbb{R},\end{equation}  where
\begin{equation}
J^{(x)} (x,\lambda)=\left(
                     \begin{array}{cc}
                       1 & \frac{b(\lambda)}{\overline{a(\bar{\lambda})}}e^{-2i\lambda^2x} \\
                      -\frac{\overline{b(\bar{\lambda})}}{a(\lambda)}e^{2i\lambda^2x} & \frac{1}{a(\lambda)\overline{a(\bar{\lambda})}} \\
                     \end{array}
                   \right),\,\,\,\lambda^2\in \mathbb{R}.
\end{equation}
  and $M^{(x)}(x,\lambda)$ satisfy asymptotic properties $M^{(x)} (x,\lambda)=I+O(\frac{1}{\lambda}),\,\,\,\,\lambda\rightarrow\infty.$
  \item We assume that the spectral function $a(\lambda)$ has $2\tilde{\wedge}$ simple zeros($2\tilde{\wedge}=2\tilde{\wedge}_1+2\tilde{\wedge}_2$), such that, $\tilde{\zeta}_j$($j=1,2,\cdots,2\tilde{\wedge}_1$) lie in $D_1\bigcup D_2\bigcup D_3$, $\bar{\tilde{\zeta}}_j$($j=2\tilde{\wedge}_1+1,2\tilde{\wedge}_1+2,\cdots,2\tilde{\wedge}$) lie in $D_4\bigcup D_5\bigcup D_6$.
   The first column of $M_{-}^{(x)}(x,\lambda)$ has simple poles at $\lambda=\bar{\tilde{\zeta}}_j$, $j=1,2,\cdots,2\tilde{\wedge}_2$, the second column of $M_{+}^{(x)}(x,\lambda)$ has simple poles at $\lambda=\tilde{\zeta}_j$, $j=1,2,\cdots,2\tilde{\wedge}_1$.
The associated residues are given by
\begin{equation} Res \{[M^{(x)}(x,\lambda)]_{2} ,  \tilde{\zeta}_j \}=\frac{b(\tilde{\zeta}_j)}{\dot{a}(\tilde{\zeta}_j)}e^{-2i\tilde{\zeta}_j^2x}[M^{(x)}(x,\tilde{\zeta}_j)]_{1}, j=1,2,\cdots,2\tilde{\wedge}_1.\end{equation}
\begin{equation} Res \{[M^{(x)}(x,\lambda)]_{1} ,  \bar{\tilde{\zeta}}_j \}=\frac{\overline{b(\tilde{\zeta}_j)}}{\overline{\dot{a}(\tilde{\zeta}_j)}}e^{2i\bar{\tilde{\zeta}}_j^2x}[M^{(x)}(x,\bar{\tilde{\zeta}}_j)]_{2} , j=2\tilde{\wedge}_1+1,2\tilde{\wedge}_1+2,\cdots,2\tilde{\wedge}.\end{equation}
\end{itemize}
\textbf{Proof}.
 We note that $M_+^{(x)}(x,\lambda)$ and $M_-^{(x)}(x,\lambda)$ have the following forms,
\begin{equation}
\begin{array}{ll}
M_+^{(x)}(x,\lambda)=(\mu_3^{D_1\cup D_2\cup D_3}(x,\lambda),\frac{\mu_1^{D_1\cup D_2\cup D_3}(x,\lambda)}{a(\lambda)})\\
M_-^{(x)}(x,\lambda)=(\frac{\mu_1^{D_4\cup D_5\cup D_6}(x,\lambda)}{\overline{a(\bar{\lambda})}},\mu_3^{D_4\cup D_5\cup D_6}(x,\lambda))
\end{array}\end{equation}
According to Eq.(2.52), we only need to set $t=0$,
\begin{equation}\left\{
  \begin{array}{ll}
   \overline{a(\bar{\lambda})}\mu_3^{D_1\cup D_2\cup D_3}(x,\lambda)+\overline{b(\bar{\lambda})}e^{2i\lambda^2x}\mu_3^{D_4\cup D_5\cup D_6}(x,\lambda)=\mu_1^{D_4\cup D_5\cup D_6}(x,\lambda)  \\
   b(\lambda)e^{-2i\lambda^2x}\mu_3^{D_1\cup D_2\cup D_3}(x,\lambda)+a(\lambda)\mu_3^{D_4\cup D_5\cup D_6}(x,\lambda)=\mu_1^{D_1\cup D_2\cup D_3}(x,\lambda)
  \end{array}
\right.\end{equation}
by direct calculation, we can  derive that the jump matrices $J^{(x)}(x,\lambda)$(Eq.(4.5)) satisfy the jump condition Eq.(4.4).\\
Consider the second column of $M_+^{(x)}(x,\lambda)$, the simple zeros $\tilde{\zeta}_j$ ($j=1,2,\cdots,2\tilde{\wedge}_1$) of $a(\lambda)$ are the simple poles of $\frac{\mu_1^{D_1\cup D_2\cup D_3}(x,\lambda)}{a(\lambda)}$. Then we have\\
\begin{equation}\begin{array}{cc}
Res\{\frac{\mu_1^{D_1\cup D_2\cup D_3}(x,\lambda)}{a(\lambda)},\tilde{\zeta}_j\}=\lim_{\lambda\rightarrow\tilde{\zeta}_j}(\lambda-\tilde{\zeta}_j)\frac{\mu_1^{D_1\cup D_2\cup D_3}(x,\lambda)}{a(\lambda)}=\frac{\mu_1^{D_1\cup D_2\cup D_3}(x,\tilde{\zeta}_j)}{\dot{a}(\tilde{\zeta}_j)},\,\,\,\, j=1,2,\cdots,2\tilde{\wedge}_1.\end{array}\end{equation}
Taking $\lambda=\tilde{\zeta}_j$ into the second equation of Eq.(4.8) we obtain
\begin{equation}\begin{array}{cc}
 \mu_1^{D_1\cup D_2\cup D_3}(x,\tilde{\zeta}_j)=[e^{-2i\lambda^2x}b(\lambda)\mu_3^{D_1\cup D_2\cup D_3}(x,\lambda)+a(\lambda)\mu_3^{D_4\cup D_5\cup D_6}(x,\lambda)]_{\lambda=\tilde{\zeta}_j} \\
= e^{-2i\tilde{\zeta}_j^2x}b(\tilde{\zeta}_j)\mu_3^{D_1\cup D_2\cup D_3}(x,\tilde{\zeta}_j),\,\,\,\, j=1,2,\cdots,2\tilde{\wedge}_1.
\end{array}\end{equation}
Furthermore,
\begin{equation}
\begin{array}{ll}
 Res \{[M_+^{(x)}(x,\lambda)]_{2} ,  \tilde{\zeta}_j \}=Res\{\frac{\mu_1^{D_1\cup D_2\cup D_3}(x,\lambda)}{a(\lambda)},\tilde{\zeta}_j\}=\frac{e^{-2i\tilde{\zeta}_j^2x}b(\tilde{\zeta}_j)\mu_3^{D_1\cup D_2\cup D_3}(x,\tilde{\zeta}_j)}{\dot{a}(\tilde{\zeta}_j)}\\
=\frac{b(\tilde{\zeta}_j)}{\dot{a}(\tilde{\zeta}_j)}e^{-2i\tilde{\zeta}_j^2x}[M_+^{(x)}(x,\tilde{\zeta}_j)]_{1},\,\,\,\, j=1,2,\cdots,2\tilde{\wedge}_1.
\end{array}
\end{equation}
It is equivalent to Eq.(4.6) and using the same proof method we can derive out Eq.(4.7).\\
\textbf{4.2.The spectral functions $\{A(\lambda),B(\lambda)\}$ }\\
\textbf{Definition 4.2.1.}  Let $g_{0}(y)$, $g_{1}(y)$ and $g_{2}(y)$ be smooth functions,
we define the map
\begin{equation}
\mathbb{\tilde{S}}: \{g_0(y),g_1(y),g_{2}(y)\}\rightarrow \{A(\lambda),B(\lambda) \}
\end{equation}
by
\begin{equation}
\left(
  \begin{array}{c}
    B(\lambda) \\
    A(\lambda) \\
  \end{array}
\right)=\mu_2^{(2)}(0,y;\lambda)=\left(
  \begin{array}{c}
    \mu_2^{12}(0,y;\lambda) \\
    \mu_2^{22}(0,y;\lambda) \\
  \end{array}
\right),\,\,\,  Im\lambda^6> 0,
\end{equation}
where $\mu_2(0,y;\lambda)$ is the unique solution of the Volterra linear integral equation
$$\mu_2(0,y;\lambda)=I+\int_{T}^{y}e^{2i\lambda^6(y-\eta)\hat{\sigma}_3}(N_2\mu_2)(0,\eta,\lambda)d\eta$$
and $N_2(0, y;\lambda)$ is given by Eq.(2.42). The spectral functions $A(\lambda)$ and $B(\lambda)$ have the
same properties with \textbf{Proposition 2.4.4}.\\
\textbf{Definition 4.2.2.} We define $\mathbb{\tilde{Q}}$ is the inverse map of $\mathbb{\tilde{S}}$,
\begin{equation}
\mathbb{\tilde{Q}}: \{A(\lambda),B(\lambda) \}\rightarrow \{g_0(y),g_1(y),g_2(y)\}
\end{equation}
where
\begin{equation}g_0(y)=2im_{12}^{(1)}(y)e^{2i\int_{0}^{y}\Delta_2(\eta)d\eta}\end{equation}
\begin{equation}g_1(y)=(4m_{12}^{(3)}(y)-2g_0(y)m_{12}^{(1)}(y))e^{2i\int_{0}^{y}\Delta_2(\eta)d\eta}-ig_0(y)(2m_{22}^{(2)}(y)+g_0(y))\end{equation}
\begin{equation}g_2(y)=[8im_{12}^{(5)}+4ig_0(y)m_{12}^{(3)}+2(g_1(y)-ig_0^2(y))m_{12}^{(1)}]e^{2i\int_{0}^{y}\Delta_2(\eta)d\eta}-g_0(y)(4m_{22}^{(4)}(y)+3ig_1(y)+g_0^2(y))   \end{equation}
where the functions $m^{(j)}(y)(j=1,2,3,4,5.)$ are determined by the asymptotic expansion Eq.(4.21), and $M^{(y)}(y,\lambda)$ is the unique solution of the RHP in the following \textbf{Theorem 4.2.3}.

\textbf{Theorem 4.2.3}
\begin{itemize}
  \item $M^{(y)}(y,\lambda)$ is the unique solution of the following RHP, \\
$
M^{(y)}(y,\lambda)=\left\{
                             \begin{array}{ll}
                              M_{-}^{(y)} (y,\lambda), & Im \lambda^6 <0 \\
                              M_{+}^{(y)} (y,\lambda), & Im \lambda^6 >0
                             \end{array}
                           \right.$ is a sectionally meromorphic function, then $M^{(y)}(y,\lambda)$ satisfied the jump condition
\begin{equation} M_{+}^{(y)} (y,\lambda)=M_{-}^{(y)} (y,\lambda)J^{(y)} (y,\lambda),\,\,\,\,\lambda^6\in \mathbb{R} \end{equation}
where,
\begin{equation}
J^{(y)} (y,\lambda)=\left(
                     \begin{array}{cc}
                       \frac{1}{A(\lambda)\overline{A(\bar{\lambda})}} & \frac{B(\lambda)}{\overline{A(\bar{\lambda})}}e^{-4i\lambda^6y} \\
                      -\frac{\overline{B(\bar{\lambda})}}{A(\lambda)}e^{4i\lambda^6y} & 1\\
                     \end{array}
                   \right),\,\,\,\lambda^6\in \mathbb{R}.
\end{equation}
and $M^{(y)}(y,\lambda)$ satisfy asymptotic properties\begin{equation} M^{(y)} (y,\lambda)=I+\frac{m^{(1)}}{\lambda}+\frac{m^{(2)}}{\lambda^2}+\frac{m^{(3)}}{\lambda^3}+\frac{m^{(4)}}{\lambda^4}+\frac{m^{(5)}}{\lambda^5}+O(\frac{1}{\lambda^6}),\,\,\,\,\lambda\rightarrow\infty.\end{equation}
  \item We assume that the spectral function $A(\lambda)$ has $2k$ simple zeros($2k=2k_1+2k_2$), such that, $\gamma_j$($j=1,2,\cdots,2k_1$) lie in $D_1\bigcup D_3\bigcup D_5$, $\bar{\gamma}_j$($j=2k_1+1,2k_1+2,\cdots,2k$) lie in $D_2\bigcup D_4\bigcup D_6$.
   The first column of $M_{+}^{(y)}(y,\lambda)$ has simple poles at $\lambda=\gamma_j$, $j=1,2,\cdots,2k_1$, the second column of $M_{-}^{(y)}(y,\lambda)$ has simple poles at $\lambda=\bar{\gamma}_j$, $j=1,2,\cdots,2k_2$.
Then, the associated residues are given by
\begin{equation} Res \{[M^{(y)}(y,\lambda)]_{1} ,  \gamma_j \}=\frac{1}{\dot{A}(\gamma_j)B(\zeta_j)}e^{4i\gamma_j^6y}[M^{(y)}(y,\zeta_j)]_{2}, j=1,2,\cdots,2k_1.\end{equation}
\begin{equation} Res \{[M^{(y)}(y,\lambda)]_{2} ,  \bar{\gamma_j} \}=\frac{1}{\overline{\dot{A}(\gamma_j)}\overline{B(\gamma_j)}}e^{-4i\bar{\zeta_j}^6y}[M^{(y)}(y,\bar{\gamma_j})]_{2} , j=2k_1+1,2k_1+2,\cdots,2k.\end{equation}

\end{itemize}
\textbf{Proof}. We set
\begin{equation}
\begin{array}{ll}
M_+^{(y)}(y,\lambda)=(\frac{\mu_3^{D_1\cup D_3\cup D_5}(y,\lambda)}{A(\lambda)},\mu_2^{D_1\cup D_3\cup D_5}(y,\lambda)),\,\,\,\lambda\in D_1\cup D_3\cup D_5 \\
M_-^{(y)}(y,\lambda)=(\mu_2^{D_2\cup D_4\cup D_6}(y,\lambda),\frac{\mu_3^{D_2\cup D_4\cup D_6}(y,\lambda)}{\overline{A(\bar{\lambda})}}),\,\,\,\lambda\in D_2\cup D_4\cup D_6.
\end{array}\end{equation}
In order to obtain the jump matrix $J^{(y)}(y,\lambda)$, we set $x=0$ according to Eq.(2.52),
\begin{equation}\left\{
  \begin{array}{ll}
   \overline{A(\bar{\lambda})}\mu_3^{D_1\cup D_3\cup D_5}(y,\lambda)+\overline{B(\bar{\lambda})}e^{4i\lambda^6y}\mu_3^{D_2\cup D_4\cup D_6}(y,\lambda)=\mu_2^{D_2\cup D_4\cup D_6}(y,\lambda)  \\
   B(\lambda)e^{-4i\lambda^6y}\mu_3^{D_1\cup D_3\cup D_5}(y,\lambda)+A(\lambda)\mu_3^{D_2\cup D_4\cup D_6}(y,\lambda)=\mu_2^{D_1\cup D_3\cup D_5}(y,\lambda)
  \end{array}
\right.\end{equation}
by direct calculation, we can  derive that the jump matrices $J^{(y)}(y,\lambda)$(Eq.(4.20)) satisfy the jump condition.\\
Consider the first column of $M_+^{(x)}(x,\lambda)$, the simple zeros $\gamma_j$ ($j=1,2,\cdots,2k_1$) of $A(\lambda)$ are the simple poles of $\frac{\mu_3^{D_1\cup D_3\cup D_5}(y,\lambda)}{A(\lambda)}$. Then we have\\
\begin{equation}\begin{array}{cc}
Res\{\frac{\mu_3^{D_1\cup D_3\cup D_3}(y,\lambda)}{A(\lambda)},\gamma_j\}=\lim_{\lambda\rightarrow\gamma_j}(\lambda-\gamma_j)\frac{\mu_3^{D_1\cup D_3\cup D_5}(y,\lambda)}{A(\lambda)}=\frac{\mu_3^{D_1\cup D_2\cup D_3}(y,\zeta_j)}{\dot{A}(\zeta_j)},\,\,\,\, j=1,2,\cdots,2k_1.\end{array}\end{equation}
Taking $\lambda=\gamma_j$ into the second equation of Eq.(4.25) we obtain
\begin{equation}\begin{array}{cc}
 \mu_3^{D_1\cup D_3\cup D_5}(y,\gamma_j)=\frac{e^{4i\lambda^6y}}{B(\lambda)}[\mu_2^{D_1\cup D_3\cup D_3}(y,\lambda)-A(\lambda)\mu_3^{D_2\cup D_4\cup D_6}(y,\lambda)]_{\lambda=\gamma_j} \\
= \frac{e^{4i\zeta_j^6y}}{B(\gamma_j)}\mu_2^{D_1\cup D_3\cup D_5}(y,\gamma_j),\,\,\,\, j=1,2,\cdots,2k_1.
\end{array}\end{equation}
Furthermore,
\begin{equation}
\begin{array}{ll}
 Res \{[M_+^{(y)}(y,\lambda)]_{1} ,  \gamma_j \}=Res\{\frac{\mu_3^{D_1\cup D_3\cup D_5}(y,\lambda)}{A(\lambda)},\gamma_j\}=\frac{e^{4i\gamma_j^6y}\mu_2^{D_1\cup D_3\cup D_5}(y,\gamma_j)}{B(\zeta_j)\dot{A}(\zeta_j)}\\
=\frac{e^{4i\gamma_j^6y}}{B(\gamma_j)\dot{A}(\gamma_j)}[M_+^{(y)}(y,\gamma_j)]_{2},\,\,\,\, j=1,2,\cdots,2k_1.
\end{array}
\end{equation}
It is equivalent to Eq.(4.22) and using the same proof method we can derive out Eq.(4.23).\\
\textbf{4.3.The spectral functions $\{\alpha(\lambda),\beta(\lambda)\}$ }\\
\textbf{Definition 4.3.1.}Given the spectral functions
\begin{equation} S_3(\lambda)=S_1^{-1}(\lambda)S_2(\lambda)=\left(
                               \begin{array}{cc}
                                 \overline{\alpha(\bar{\lambda})} & \beta(\lambda) \\
                                 \overline{\beta(\bar{\lambda})} & \alpha(\lambda) \\
                               \end{array}
                             \right),
\end{equation}
and the smooth functions $h_L(x)=u(x,L)$.
We define the map
\begin{equation} \mathbb{\tilde{\tilde{S}}}: \{h_T(x)\}\rightarrow \{\alpha(\lambda),\beta(\lambda) \} \end{equation}
by
\begin{equation}\mu_2(x,L,\lambda)=\mu_1(x,L,\lambda)e^{-i(\lambda^2x+2\lambda^6L)\hat{\sigma}_3}S_3(\lambda)\end{equation}
where $\mu_1(x,L;\lambda), \mu_2(x,L,\lambda)$ is the unique solution of the Volterra linear integral equation
$\mu_2(x,L;\lambda)=I+\int_{0}^{x}e^{-i\lambda^2(x-\xi)\hat{\sigma}_3}(N_1\mu_2)(\xi,L;\lambda)d\xi$ and
$\mu_1(x,L;\lambda)=I-\int_{x}^{\infty}e^{-i\lambda^2(x-\xi)\hat{\sigma}_3}(N_1\mu_1)(\xi,L;\lambda)d\xi$,  $N_1(x, L;\lambda)$ is given by
\[
N_1(x,L;\lambda)=\left(
                   \begin{array}{cc}
                     -ih_L(x) & \lambda h_L(x)e^{-\int_{0}^{x}ih_L(\xi,L)d\xi} \\
                     2\lambda e^{\int_{0}^{x}ih_L(\xi,L)d\xi}  &  ih_L(x)\\
                   \end{array}
                 \right).
\]
\textbf{Proposition 4.3.2.} The spectral functions $\alpha(\lambda)$ and $\beta(\lambda)$ have the
following properties
\begin{description}
  \item[(i)] $\alpha(\lambda)$ and $\beta(\lambda)$ are analytic for $Im\lambda^2<0$ and continuous and bounded for $Im\lambda^2\leq 0$;
  \item[(ii)]$\alpha(\lambda)=1+\sum_{j=1}^m \frac{\alpha_j(\lambda)}{\lambda^m}+O(\frac{1}{\lambda^{m+1}}),\,\,\,\, \beta(\lambda)=\sum_{j=1}^m \frac{\beta_j(\lambda)}{\lambda^m}+O(\frac{1}{\lambda^{m+1}})$ as $\lambda\rightarrow\infty$, $Im\lambda^2\leq 0$;
  \item[(iii)] $\alpha(\lambda)\overline{\alpha(\bar{\lambda})}-\beta(\lambda)\overline{\beta(\bar{\lambda})}=1$, $\lambda^2\in \mathbb{R}$;
\item[(iv)] $\alpha(-\lambda)=\alpha(\lambda),\beta(-\lambda)=-\beta(\lambda)$, $Im\lambda^2\leq 0$.
\end{description}
\textbf{Definition 4.3.3.}
We define $\mathbb{\tilde{\tilde{Q}}}$ is the inverse map of $\mathbb{\tilde{\tilde{S}}}$,
\begin{equation}
\mathbb{\tilde{\tilde{Q}}}: \{\alpha(\lambda),\beta(\lambda) \}\rightarrow \{h_L(x)\}
\end{equation} where
$
h_L(x)=2im_L(x)e^{2i\int_{0}^{x}\Delta_1(\xi,L)d\xi},
m_L(x)=\lim_{\lambda\rightarrow\infty}(\lambda M^{(L)}(x,\lambda))_{12}.\\
$
\textbf{Theorem 4.3.4}
\begin{itemize}
  \item $M^{(L)}(x,\lambda)$ is the unique solution of the following RHP\\
$M^{(L)}(x,\lambda)=\left\{
                             \begin{array}{ll}
                              M_{-}^{(L)} (x,\lambda), & Im \lambda^2 <0 \\
                              M_{+}^{(L)} (x,\lambda), & Im \lambda^2 >0
                             \end{array}
                           \right.$ is a sectionally meromorphic function, then $M^{(L)}(x,\lambda)$ satisfied the jump condition
 \begin{equation}M_{+}^{(L)} (x,\lambda)=M_{-}^{(L)} (x,\lambda)J^{(L)} (x,\lambda),\,\,\,\lambda^2\in \mathbb{R},\end{equation}  where
\begin{equation}
J^{(L)} (x,\lambda)=\left(
                     \begin{array}{cc}
                       1 &- \frac{\beta(\lambda)}{\overline{\alpha(\bar{\lambda})}}e^{-2i(\lambda^2x+2\lambda^6L)} \\
                      \frac{\overline{\beta(\bar{\lambda})}}{\alpha(\lambda)}e^{2i(\lambda^2x+2\lambda^6L)} & \frac{1}{\alpha(\lambda)\overline{\alpha(\bar{\lambda})}} \\
                     \end{array}
                   \right),\,\,\,\lambda^2\in \mathbb{R}.
\end{equation}
  and $M^{(L)}(x,\lambda)$ satisfy asymptotic properties $M^{(L)} (x,\lambda)=I+O(\frac{1}{\lambda}),\,\,\,\,\lambda\rightarrow\infty.$
  \item We assume that the spectral function $\alpha(\lambda)$ has $2\tilde{\vee}$ simple zeros($2\tilde{\vee}=2\tilde{\vee}_1+2\tilde{\vee}_2$), such that, $\tilde{\varepsilon}_j$($j=1,2,\cdots,2\tilde{\vee}_1$) lie in $D_4\bigcup D_4\bigcup D_6$, $\bar{\tilde{\varepsilon}}_j$($j=2\tilde{\vee}_1+1,2\tilde{\vee}_1+2,\cdots,2\tilde{\vee}$) lie in $D_1\bigcup D_2\bigcup D_3$.
   The second column of $M_{-}^{(L)}(x,\lambda)$ has simple poles at $\lambda=\bar{\tilde{\varepsilon}}_j$, $j=1,2,\cdots,2\tilde{\vee}_2$, the first column of $M_{+}^{(L)}(x,\lambda)$ has simple poles at $\lambda=\tilde{\varepsilon}_j$, $j=1,2,\cdots,2\tilde{\vee}_1$.
The associated residues are given by
\begin{equation} Res \{[M^{(L)}(x,\lambda)]_{1} ,  \tilde{\varepsilon}_j \}=\frac{e^{2i(\tilde{\varepsilon}_j^2x+2\tilde{\varepsilon}_j^6L)}}{\dot{\alpha}(\tilde{\varepsilon}_j)\beta(\tilde{\varepsilon}_j)}[M^{(L)}(x,\tilde{\varepsilon}_j)]_{2}, j=1,2,\cdots,2\tilde{\vee}_1.\end{equation}
\begin{equation} Res \{[M^{(L)}(x,\lambda)]_{2} ,  \bar{\tilde{\varepsilon}}_j \}=\frac{e^{-2i(\bar{\tilde{\varepsilon}}_j^2x+2\bar{\tilde{\varepsilon}}_j^6L)}}{\overline{\beta(\tilde{\varepsilon}_j)}\overline{\dot{\alpha}(\tilde{\varepsilon}_j)}}[M^{(L)}(x,\bar{\tilde{\varepsilon}}_j)]_{1} , j=2\tilde{\vee}_1+1,2\tilde{\vee}_1+2,\cdots,2\tilde{\vee}.\end{equation}

\end{itemize}
\textbf{Proof}.
 We note that $M_+^{(L)}(x,\lambda)$ and $M_-^{(L)}(x,\lambda)$ have the following forms,
\begin{equation}
\begin{array}{ll}
M_-^{(L)}(x,\lambda)=(\frac{\mu_1^{D_4\cup D_5\cup D_6}(x,L;\lambda)}{\alpha(\lambda)} ,\mu_2^{D_4\cup D_5\cup D_6}(x,L;\lambda))\\
M_+^{(L)}(x,\lambda)=(\mu_2^{D_1\cup D_2\cup D_3}(x,L;\lambda),\frac{\mu_1^{D_1\cup D_2\cup D_3}(x,L;\lambda)}{\overline{\alpha(\bar{\lambda})}},)
\end{array}\end{equation}
According to Eq.(2.54), we only need to set $y=L$,
\begin{equation}\left\{
  \begin{array}{ll}
   \overline{\alpha(\bar{\lambda})}\mu_1^{D_4\cup D_5\cup D_6}(x,L;\lambda)+\overline{\beta(\bar{\lambda})}e^{2i(\lambda^2x+2\lambda^6L)}\mu_1^{D_1\cup D_2\cup D_3}(x,L;\lambda)=\mu_2^{D_1\cup D_2\cup D_3}(x,L;\lambda)  \\
   \beta(\lambda)e^{-2i(\lambda^2x+2\lambda^6L)}\mu_1^{D_4\cup D_5\cup D_6}(x,L;\lambda)+\alpha(\lambda)\mu_1^{D_1\cup D_2\cup D_3}(x,L;\lambda)=\mu_2^{D_4\cup D_5\cup D_6}(x,L;\lambda)
  \end{array}
\right.\end{equation}
by direct calculation, we can  derive that the jump matrices $J^{(L)}(x,\lambda)$(Eq.(4.34)) satisfy the jump condition Eq.(4.33).\\
Consider the frist column of $M_-^{(L)}(x,\lambda)$, the simple zeros $\tilde{\varepsilon}_j$ ($j=1,2,\cdots,2\tilde{\vee}_1$) of $\alpha(\lambda)$ are the simple poles of $\frac{\mu_1^{D_4\cup D_5\cup D_6}(x,L;\lambda)}{\alpha(\lambda)}$. Then we have\\
\begin{equation}\begin{array}{cc}
Res\{\frac{\mu_1^{D_4\cup D_5\cup D_6}(x,L;\lambda)}{\alpha(\lambda)},\tilde{\varepsilon}_j\}=\lim_{\lambda\rightarrow\tilde{\varepsilon}_j}(\lambda-\tilde{\varepsilon}_j)\frac{\mu_1^{D_4\cup D_5\cup D_6}(x,L;\lambda)}{\alpha(\lambda)}=\frac{\mu_1^{D_4\cup D_5\cup D_6}(x,\tilde{\varepsilon}_j)}{\dot{\alpha}(\tilde{\varepsilon}_j)},\,\,\,\, j=1,2,\cdots,2\tilde{\vee}_1.\end{array}\end{equation}
Taking $\lambda=\tilde{\varepsilon}_j$ into the second equation of Eq.(4.38) we obtain
\begin{equation}\begin{array}{cc}
 \mu_1^{D_4\cup D_5\cup D_6}(x,\tilde{\varepsilon}_j)=\frac{e^{2i(\tilde{\varepsilon}_j^2x+2\tilde{\varepsilon}_j^6L)}}{\beta(\tilde{\varepsilon}_j)}[\mu_2^{D_4\cup D_5\cup D_6}(x,\lambda)-\alpha(\lambda)\mu_1^{D_1\cup D_2\cup D_3}(x,\lambda)]_{\lambda=\tilde{\varepsilon}_j} \\
= \frac{e^{2i(\tilde{\varepsilon}_j^2x+2\tilde{\varepsilon}_j^6L)}}{\beta(\tilde{\varepsilon}_j)}\mu_2^{D_4\cup D_5\cup D_6}(x,\tilde{\varepsilon}_j),\,\,\,\, j=1,2,\cdots,2\tilde{\vee}_1.
\end{array}\end{equation}
Furthermore,
\begin{equation}
\begin{array}{ll}
 Res \{[M_-^{(L)}(x,\lambda)]_{1} ,  \tilde{\varepsilon}_j \}=Res\{\frac{\mu_1^{D_4\cup D_5\cup D_6}(x,\lambda)}{\alpha(\lambda)},\tilde{\varepsilon}_j\}=\frac{e^{2i(\tilde{\varepsilon}_j^2x+2\tilde{\varepsilon}_j^6L)}}{\dot{\alpha}(\tilde{\varepsilon}_j)\beta(\tilde{\varepsilon}_j)}\mu_2^{D_4\cup D_5\cup D_6}(x,\tilde{\varepsilon}_j)\\
=\frac{e^{2i(\tilde{\varepsilon}_j^2x+2\tilde{\varepsilon}_j^6L)}}{\dot{\alpha}(\tilde{\varepsilon}_j)\beta(\tilde{\varepsilon}_j)}[M_-^{(L)}(x,\tilde{\varepsilon}_j)]_{2},\,\,\,\, j=1,2,\cdots,2\tilde{\vee}_1.
\end{array}
\end{equation}
It is equivalent to Eq.(4.35) and using the same proof method we can derive out Eq.(4.36).\\
\textbf{4.4. The principal RHP}\\
\textbf{Theorem 4.4.1.} Let $u_0(x)\in S(\mathbb{R^{+}})$ is a smooth function. Suppose that the function $g_0(y),g_1(y),g_2(y)$ are compatible with the function $u_0(x)$ at $x=y=0$. Define the spectral function $a(\lambda)$, $b(\lambda)$, $A(\lambda)$ and $B(\lambda)$, in terms of  $u_0(x)$, $g_0(y),g_1(y)$, and $g_2(y)$ of \textbf{Definition 4.1.2} and \textbf{Definition 4.2.2}. Suppose that $a(\lambda)$, $b(\lambda)$, $A(\lambda)$ and $B(\lambda)$
satisfy the global relation
\[a(\lambda)B(\lambda)-b(\lambda)A(\lambda)=e^{4i\lambda^6L}c^+(\lambda),\,\,\,Im \lambda^6 \geq 0,\]
where $S_1(\lambda)=\mu_1(0,0;\lambda),S_2(\lambda)=S_2(L,\lambda)=(e^{2i\lambda^6L}\mu_3(0,L;\lambda))^{-1}$, if $\lambda \rightarrow \infty$ the global relation is replaced by $a(\lambda)B(\lambda)-b(\lambda)A(\lambda)=0$.
Assume that the possible zeros of $\{\zeta_j\}_{j=1}^{2\wedge}$ are $a(\lambda)$ and $\{\varepsilon_j\}_{j=1}^{2\vee}$ of $\alpha(\lambda)$, then define the $M(x,y,\lambda)$ as the solution of the following RHP
\begin{itemize}
  \item $M(x,y;\lambda)$ is sectionally meromorphic in $\mathbb{C}\setminus \{ \lambda^{6} \in \mathbb{R}\}$.
  \item The first column of $M(x,y;\lambda)$ has simple poles at $\lambda=\bar{\zeta}_j$, $j=1,2,\cdots,2\wedge_2$, and $\lambda=\varepsilon_j$, $j=1,2,\cdots,2\vee_1$. The second column of $M(x,y;\lambda)$ has simple poles at $\lambda=\zeta_j$, $j=1,2,\cdots,2\wedge_1$, and $\lambda=\bar{\varepsilon}_j$, $j=1,2,\cdots,2\vee_2$.
  \item $M(x,y;\lambda)$ satisfies the jump condition
\begin{equation}
M_{+}(x,y;\lambda)=M_{-}(x,y;\lambda)J(x,y;\lambda),\,\,\,\lambda^{6} \in \mathbb{R}.
\end{equation}
\item $M(x,y;\lambda)=I+\mathcal{O}(\frac{1}{\lambda}),\,\,\,\lambda\rightarrow\infty$.
\item $M(x,y;\lambda)$ satisfies the residue conditions of \textbf{Proposition2.7.2}.
\end{itemize}
Then $M(x,y;\lambda)$ exists and is unique, we define $u(x,y)$ in terms of  $M(x,y;\lambda)$ by
\begin{equation}\begin{array}{cc}
u(x,y)=2im(x,y)e^{2i\int_{(0,0)}^{(x,y)}\Delta},\\
m(x,y)=\lim_{\lambda\rightarrow\infty}(\lambda M(x,y;\lambda))_{12}.
\end{array}\end{equation}
Furthermore $u(x,y)$ is the solution of the cSTO equation (1), and $u(x,0)=u_0(x)$, $u(0,t)=g_0(t)$, $u_x(0,y)=g_1(y)$, $u_{xx}(0,y)=g_2(y)$.\\
\textbf{Proof.} In fact, if we assume that $a(\lambda)$ and $\alpha(\lambda)$ have no zeroes, then the $2\times2$ function $M(x,y;\lambda)$ satisfies a non-sigular RHP. Using the fact that the jump matrix $J(x,y;\lambda)$
matches with the symmetry conditions, we can show that this problem has a unique global solution [26]. The case that $a(\lambda)$ and $\alpha(\lambda)$ have a finite number of zeros can be mapped to the case of no zeros supplemented by an algebraic system of equations which is always uniquely solvable.\\
\textbf{Theorem 4.2.2.} The RHP in\textbf{Theorem 4.2.1} with the vanishing boundary condition $M(x,y;\lambda)\rightarrow 0$ ($\lambda \rightarrow \infty)$, has only the zero solution.\\
\textbf{Proof.} Assume that $M(x,y;\lambda)$ is a solution of the RHP in \textbf{Theorem 4.2.1} such that $M_{\pm}(x,y;\lambda)\rightarrow\infty$ ($\lambda \rightarrow \infty)$. $A$ is a $2\times2$ matrix, $A^{\dag}$ denotes
the complex conjugate transpose of $A$.\\ Define
\begin{equation}\begin{array}{cc}
\chi_{+}(\lambda)=M_+(\lambda)M_{-}^{\dag}(-\bar{\lambda}),\,\,\,Im\lambda^6\geq0,\\
\chi_{-}(\lambda)=M_-(\lambda)M_{+}^{\dag}(-\bar{\lambda}),\,\,\,Im\lambda^6\leq0,
\end{array}\end{equation}
where the $x$ and $y$ are dependence. $\chi_{+}(\lambda)$ and  $\chi_{-}(\lambda)$ are analytic in $\{\lambda\in \mathbb{C} \setminus Im\lambda^6>0\}$ and $\{\lambda\in \mathbb{C} \setminus Im\lambda^6<0\}$
respectively. By the symmetry relations $a(-\lambda)=a(\lambda),b(-\lambda)=-b(\lambda),$ and $A(-\lambda)=A(\lambda),B(-\lambda)=-B(\lambda)$ and Eq.(31), we infer that
\begin{equation}
J_{1}^{\dag}(-\bar{\lambda})=J_1(\lambda),\,\,\,\,J_{2}^{\dag}(-\bar{\lambda})=J_2(\lambda),\,\,\,\,J_{3}^{\dag}(-\bar{\lambda})=J_3(\lambda).
\end{equation}
Then
\begin{equation}
\begin{array}{cc}
\chi_{+}(\lambda)=M_-(\lambda)J(\lambda)M_{-}^{\dag}(-\bar{\lambda}),\,\,\,Im\lambda^6\in \mathbb{R},\\
\chi_{-}(\lambda)=M_-(\lambda)J^{\dag}(-\bar{\lambda})M_{-}^{\dag}(-\bar{\lambda}),\,\,\,Im\lambda^6\in \mathbb{R}.
\end{array}
\end{equation}
Eq.(4.45) and Eq.(4.46) mean that $\chi_{+}(\lambda)=\chi_{-}(\lambda)$ for $Im\lambda^6\in \mathbb{R}$. Therefore, $\chi_{+}(\lambda)$ and $\chi_{-}(\lambda)$ define an entire function vanishing at infinity, so $\chi_{+}(\lambda)$ and $\chi_{-}(\lambda)$ are identically zero. Noting $J_1(i\kappa)(\kappa\in\mathbb{R}),$ is a Hermitian matrix with unit determinant and $(2,2)$ entry $1$ for any $\kappa\in\mathbb{R}$. Therefore, $J_1(i\kappa)(\kappa\in\mathbb{R})$
is a positive definite matrix. Since $\chi_{-}(\kappa)$ vanishes identically
for $\kappa\in i\mathbb{R}$, i.e.
\begin{equation}
M_+(i\kappa)J_1(i\kappa)M_{+}^{\dag}(i\kappa)=0,\,\,\,\kappa\in\mathbb{R}.
\end{equation}
We can deduce that $M_+(i\kappa)=0$ as $\kappa\in\mathbb{R}$. It follows that $M_{+}(\lambda)$ and $M_-(\lambda)$ vanish identically.\\
\textbf{Proposition 4.2.3.} \textbf{Proof that $u(x, y)$ satisfies the cSTO equation.}\\
Using arguments of the dressing methods[19], it can be verified directly that if $M(x,y;\lambda)$ is
defined as the unique solution of the above RHP, and if $u(x, y)$ is defined in terms of $M(x,y;\lambda)$ by Eq.(4.43), then $u(x,y)$ and $M(x,y;\lambda)$ satisfy two parts of the Lax pair, hence $u(x,y)$
is solvable on cSTO equation.\\

\section{Acknowledgments}

This work is in part supported by the National Natural Science Foundation of China( NNSFC Grant Nos. 11271008, 61072147).

\begin{flushleft}
\textbf{Reference}
\end{flushleft}

[1] Gardner G.S., Greene, J.M., Kruskal, M.D., Miura, R.M.: Phys. Rev. Lett. 19, 1095 (1967).

[2] Faddeev L.D., Takhtajan, L.A.: Hamiltonian Methods in the Theory of Solitons. Berlin-Heidelberg-NewYork: Springer Verlag, 1987.

[3] Fokas A.S., Zakharov, V.E. (eds): Important Developments in Soliton Theory. Berlin-Heidelberg-NewYork: Springer Verlag, 1994.

[4] Fokas A.S. A unified transform method for solving linear and certain nonlinear PDEs. Proc R Soc Lond A, 1997, 453: 1411-1443.

[5] Fokas A.S, Its A R, Sung L Y. The nonlinear Schr?dinger equation on the half-line. Nonlinearity, 2005, 18:1771-1822.

[6] Its, Alexander, Shepelsky, Dmitry. Initial boundary value problem for the focusing nonlinear Schr\"{o}dinger equation with Robin boundary condition: half-line approach. Proc. R. Soc. Lond. Ser. A Math. Phys. Eng. Sci. 469 (2013), no. 2149, 20120199, 14 pp.

[7]Boutet de Monvel A, Fokas A.S, Shepelsky D. The mKDV equation on the half-line. J Inst Math Jussieu, 2004, 3: 139-164.

[8] Boutet de Monvel A, Shepelsky D. Initial boundary value problem for the MKdV equation on a finite interval. Ann Inst Fourier (Grenoble), 2004, 54: 1477-1495.

[9] Lenells Jonatan, The derivative nonlinear Schr\"{o}dinger equation on the half-line. Physica D 2008, 237: 3008每3019.

[10]Lenells Jonatan, An integrable generalization of the sine-Gordon equation on the half-line. IMA J. Appl. Math. 76 (2011), no. 4, 554-572.

[11] Lenells, Jonatan, The nonlinear steepest descent method: asymptotics for initial-boundary value problems. SIAM J. Math. Anal. 48 (2016), no. 3, 2076-2118.

[12] Xu Jian, Fan Engui, The unified transform method for the Sasa-Satsuma equation on the half-line. Proc. R. Soc. Lond. Ser. A Math. Phys. Eng. Sci.469, no.2159, 20130068, (2013)35-55.

[13] Xu Jian, Fan Engui, The three-wave equation on the half-line. Phys. Lett. A 378 (2014), no. 1-2, 26-33.

[14] Xu Jian, Fan Engui, Initial-boundary value problem for integrable nonlinear evolution equation with $3\times3$ Lax pairs on the interval. Stud. Appl. Math. 136 (2016), no. 3, 321-354.

[15] Fulton, S.R., Fokas, A.S., Xenophontos, C.A. An analytical method for linear elliptic PDEs and its numerical implementation. J. Comput. Appl. Math. 167 (2004), no. 2, 465-483.

[16] H. Tasso, Coles ansatz and extension of Burgers equation, Report No. IPP6/142 (Ber. MPI fur Plasma physik, Garching,1976).

[17] H. Tasso, Hamiltonian formulation of odd Burgers hierarchy, J. Phys. A. 29 (1996) 7779-7784.

[18] F. Verheest, W. Hereman, Nonlinear mode decoupling for classes of evolution equations, J.Phys. A. 15 (1982) 95-102.

[19] Wang, Deng-Shan; Yin, Shujuan; Tian, Ye; Liu, Yifang Integrability and bright soliton solutions to the coupled nonlinear Schr?dinger equation with higher-order effects. Appl. Math. Comput. 229 (2014), 296每309. 

[20] Dong Huan He; Guo Bao Yong; Yin Bao Shu; Generalized Fractional Supertrace Identity For Hamiltonian Structure of Nls-Mkdv Hierarchy With Self-Consistent Sources. Analysis and Mathematical Physics,6,2,199-209,2016.

[21] V. V. Gudkov, A family of exact traveling wave solutions to nonlinear evolution and wave equations, J. Phys. A 38(9),4794-4803 (1997).

[22]Chao Yue, Tiecheng Xia, Algebro-geometric solutions for the complex Sharma-Tasso-Olver hierarchy, Journal of Mathematical Physics 55, 083511 (2014).

[23] Fan Engui, A family of completely integrable multi-Hamiltonian systems explicitly related to some celebrated equations, J. Math. Phys. 42(9), 4327-4344 (2001).

[24] Fokas, A.S., Lenells, J., The unified transform for the modified Helmholtz equation in the exterior of a square. Unified transform for boundary value problems, 172每180, SIAM, Philadelphia, PA, 2015.

[25] Fokas, A.S., Lenells, J., Pelloni, B. Boundary value problems for the elliptic sine-Gordon equation in a semi-strip. J. Nonlinear Sci. 23 (2013), no. 2, 241每282.

[26] V.E.Adler, I.T.Khabibullin, Boundary conditions for integrable chains.(Russian) Funktsional. Anal.i Prilozhen. 31 (1997), no.2, 1-14, 95; translation in Funct.Anal.Appl. 31 (1997), no.2,75-85.

}
\end{document}